\documentclass[prd,showpacs,showkeys,nofootinbib,twocolumn]{revtex4-1}

\usepackage{graphicx}
\usepackage{color}
\usepackage{amsmath,amsfonts,amssymb}

\begin{document}

\def\ut{{\underset {\widetilde{\ \ }}u}}
\def\at{{\underset {\widetilde{\ \ }}a}}
\def\ap{\check{a}}

\title{Generalized deviation equation and determination of the curvature in General Relativity}

\author{Dirk Puetzfeld}
\email{dirk.puetzfeld@zarm.uni-bremen.de}
\homepage{http://puetzfeld.org}
\affiliation{ZARM, University of Bremen, Am Fallturm, 28359 Bremen, Germany} 

\author{Yuri N. Obukhov}
\email{obukhov@ibrae.ac.ru}
\affiliation{Theoretical Physics Laboratory, Nuclear Safety Institute, 
Russian Academy of Sciences, B.Tulskaya 52, 115191 Moscow, Russia} 

\date{ \today}

\begin{abstract}
We derive a generalized deviation equation -- analogous to the well-known geodesic deviation equation -- for test bodies in General Relativity. Our result encompasses and generalizes previous extensions of the standard geodesic deviation equation. We show how the standard as well as a generalized deviation equation can be used to measure the curvature of spacetime by means of a set of test bodies. In particular, we provide exact solutions for the curvature by using the standard deviation equation as well as its next order generalization.
\end{abstract}


\maketitle


\section{Introduction}\label{introduction_sec}

In General Relativity, the comparison of test bodies moving along adjacent world lines is of direct operational significance. The observation of a suitably prepared set of test bodies allows for the determination of the components of the curvature tensor. In this work we derive a deviation equation for test bodies moving along general curves in arbitrary background spacetimes. 

Our findings in this work generalize the well-known geodesic deviation equation and some of its early modifications. In contrast to many previous treatments, we are making use of a covariant expansion technique based on Synge's ``world function'' \cite{Synge:1960,DeWitt:Brehme:1960}. This expansion technique has also been applied extensively in the context of the equations of motion of extended test bodies \cite{Dixon:1964,Dixon:1974,Dixon:1979,Dixon:2008,Puetzfeld:Obukhov:2014:2,Dixon:2015,Obukhov:Puetzfeld:2015:1} and in the gravitational self-force problem \cite{Ottewill:2011,Poisson:etal:2011}.

We explicitly show, how the deviation equation can be used to measure the curvature of spacetime and thereby the gravitational field. For this we extend Szekeres' ``gravitational compass'' \cite{Szekeres:1965} and provide an exact solution for the curvature components in terms of the mutual accelerations between the constituents of a swarm of test bodies.

The structure of the paper is as follows: In section \ref{sec_gen_dev_derivation} we derive an exact generalized deviation along general world lines. Furthermore, we provide a systematic expansion of this equation in powers of the deviation vector. Subsequently we study several special cases of our general deviation equation in section \ref{sec_special_dev_derivation}. In particular, we make contact with several other suggestions for deviation equations in the literature. We then use the derived deviation equation in section \ref{sec_compass} to determine the curvature of spacetime by means of a generalized gravitational compass. We conclude our paper in section \ref{sec_conclusions} with a discussion of the results obtained and with an outlook of their possible applications. Our notations and conventions are summarized in appendix \ref{sec_notation} and table \ref{tab_symbols}. An exposition on normal coordinates given in appendix \ref{sec_normal_coordinates} contains some new results not found in the earlier literature.

\section{Generalized deviation equation}\label{sec_gen_dev_derivation}

Alternative derivations and generalizations \cite{Plebanski:1965,Hodgkinson:1972,Bazanski:1974,Hojman:1975,Bazanski:1976,Bazanski:1977:1,Bazanski:1977:2,Novello:etal:1977,Aleksandrov:Piragas:1978,Mannoff:1979,Schattner:Truemper:1981,Swaminarayan:etal:1983,Schutz:1985,Kamran:Marck:1986,Ciufolini:1986,Bazanski:etal:1987:1,Bazanski:etal:1987:2,Bazanski:1988,Vanzo:1992,Roberts:1996,Kerner:etal:2001,Mannoff:2001,Holten:2002,Chicone:Mashhoon:2002,Nieto_etal:2003,Mohseni:2004,Heydari-Fard:etal:2005,Perlick:2008,Vines:2014} as well as applications \cite{Szekeres:1965,Shirokov:1973,Greenberg:1974,Mashhoon:1975,Mashhoon:1977,Tammelo:1977,Dolan:etal:1980,Caviglia:etal:1982,Fuchs:1983,Audretsch:1983,Tammelo:1984,Bazanski:1986,Ciufolini:Demianski:1986,Bazanski:etal:1987:3,Bazanski:1989,Fuchs:1990:1,Fuchs:1990:2,Mohseni:2000,Balakin:etal:2000,Colistete:etal:2002,Biesiada:2003,Baskaran:Grishchuk:2004,Mullari:Tammelo:2006,Mohseni:2009,Bini:etal:2011,Koekoek:etal:2011} of the standard and generalized deviation equations have been extensively studied in the physical and mathematical literature. The interest in deviation equations is of course directly linked to their operational meaning, allowing for a measurement of the gravitational field, i.e.\ the curvature of spacetime, in General Relativity. 

Here we focus on a covariant derivation of a deviation equation for general curves, which are not necessarily geodesic. We base our derivation on Synge's world function $\sigma(x,y)$ \cite{Synge:1960}, which introduces a covariant generalization of the finite distance between the spacetime points $x$ and $y$. Basic definitions and our notation are summarized in appendix \ref{sec_notation}. 

\subsection{Comparison of two general curves}

Let us start by comparing two general curves $Y(t)$ and $X(\tilde{t})$ in an arbitrary spacetime manifold. At this stage even the parameters $t$ and $\tilde{t}$ can be general, i.e.\ are not necessarily the proper time on the given curves. Now we connect two points $x\in X$ and $y\in Y$ on the two curves by the geodesic joining the two points (we assume that this geodesic is unique). 

\begin{figure}
\begin{center}
\includegraphics[width=7cm]{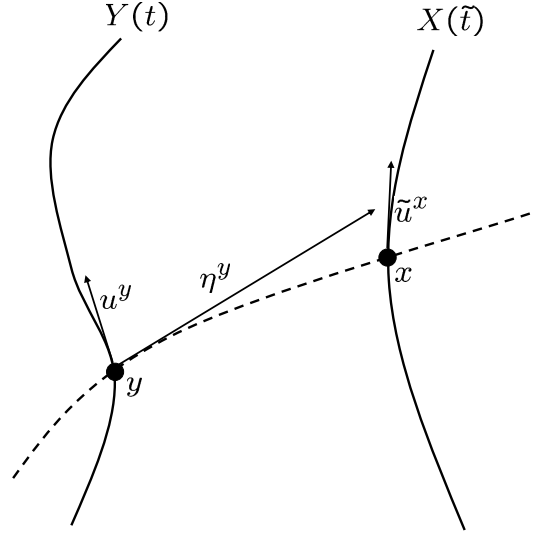}
\end{center}
\caption{\label{fig_setup} Sketch of the two arbitrarily parametrized world lines $Y(t)$ and $X(\tilde{t})$, and the geodesic connecting two points on these world line. The (generalized) deviation vector along the reference world line $Y$ is denoted by $\eta^y$.}
\end{figure}

Along the geodesic we have the world function $\sigma$, and conceptually the closest object to the connecting vector between the two points is the covariant derivative of the world function, denoted at the point $y$ by $\sigma^y$. Note though that $\sigma^y$ is just tangent at that point (its length being the the geodesic length between $y$ and $x$), only in flat spacetime it coincides with the connecting vector. Keeping in mind such an interpretation, let us now work out a propagation equation for this ``generalized'' connecting vector along the reference curve, cf.\ fig.\ \ref{fig_setup}. Following our conventions the reference curve will be $Y(t)$ and we define the generalized connecting vector to be:
\begin{eqnarray}
\eta^y := - \sigma^y \label{gen_dev_definition} 
\end{eqnarray}
Taking its covariant total derivative, we have
\begin{eqnarray}
\frac{D}{dt} \eta^{y_1} &=& - \frac{D}{dt} \sigma^{y_1}\left(Y(t),X(\tilde{t})\right) \nonumber \\
&=& - \sigma^{y_1}{}_{y_2} \frac{\partial Y^{y_2}}{\partial t} - \sigma^{y_1}{}_{x_2} \frac{\partial X^{x_2}}{\partial \tilde{t}} \frac{d\tilde{t}}{dt} \nonumber \\
&=& - \sigma^{y_1}{}_{y_2} u^{y_2} - \sigma^{y_1}{}_{x_2} \tilde{u}^{x_2} \frac{d\tilde{t}}{dt}, \label{eta_1st_deriv}
\end{eqnarray}
where in the last line we defined the velocities along the two curves $Y$ and $X$. As usual, $\sigma^y{}_{x_1\dots y_2\dots} := \nabla_{x_1}\dots\nabla_{y_2}\dots (\sigma^y)$ denote the higher order covariant derivatives of the world function. We continue by taking the second derivative of (\ref{eta_1st_deriv}), which yields
\begin{eqnarray}
\frac{D^2}{dt^2} \eta^{y_1} &=& - \sigma^{y_1}{}_{y_2 y_3} u^{y_2} u^{y_3} - 2 \sigma^{y_1}{}_{y_2 x_3} u^{y_2} \tilde{u}^{x_3} \frac{d\tilde{t}}{dt} \nonumber \\
&& - \sigma^{y_1}{}_{y_2} a^{y_2} - \sigma^{y_1}{}_{x_2 x_3} \tilde{u}^{x_2} \tilde{u}^{x_3} \left(\frac{d\tilde{t}}{dt} \right)^2 \nonumber \\
&& - \sigma^{y_1}{}_{x_2} \tilde{a}^{x_2} \left(\frac{d\tilde{t}}{dt} \right)^2 - \sigma^{y_1}{}_{x_2} \tilde{u}^{x_2}  \frac{d^2\tilde{t}}{dt^2}, \label{eta_2nd_deriv}
\end{eqnarray}
here we introduced the accelerations $a^y:={D u^y}/dt$, and $\tilde{a}^x:={D \tilde{u}^x}/d\tilde{t}$. Equation (\ref{eta_2nd_deriv}) is already the generalized deviation equation, but the goal is to have all the quantities therein defined along the reference wordline $Y$. 

We now derive some auxiliary formulas, by introducing the inverse of the second derivative of the world function via the following equations:
\begin{eqnarray}
\stackrel{-1}{\sigma}{}\!\!^{y_1}{}_x \sigma^{x}{}_{y_2} &=& \delta^{y_1}{}_{y_2}, \label{inverse_1}\\
\stackrel{-1}{\sigma}{}\!\!^{x_1}{}_y \sigma^{y}{}_{x_2} &=& \delta^{x_1}{}_{x_2}. \label{inverse_2}
\end{eqnarray}
Multiplication of (\ref{eta_1st_deriv}) by $\stackrel{-1}{\sigma}{}\!\!^{x_3}{}_{y_1}$ then yields
\begin{eqnarray}
\tilde{u}^{x_3} \frac{d\tilde{t}}{dt} &=& - \stackrel{-1}{\sigma}{}\!\!^{x_3}{}_{y_1} \sigma^{y_1}{}_{y_2} u^{y_2} + \stackrel{-1}{\sigma}{}\!\!^{x_3}{}_{y_1}  \frac{D \sigma^{y_1}}{dt} \nonumber \\ 
&=& K^{x_3}{}_{y_2} u^{y_2} - H^{x_3}{}_{y_1} \frac{D \sigma^{y_1}}{dt}. \label{formal_vel_1}
\end{eqnarray}
Where in the last line we defined two auxiliary quantities $K^x{}_{y}$ and $H^x{}_{y}$ -- the notation follows the terminology of Dixon. Equation (\ref{formal_vel_1}) allows us to formally express the the velocity along the curve $X$ in terms of the  quantities which are defined at $Y$ and then ``propagated'' by $K^x{}_{y}$ and $H^x{}_{y}$. Using (\ref{formal_vel_1}) in (\ref{eta_2nd_deriv}) we arrive at:
\begin{eqnarray}
\frac{D^2}{dt^2} \eta^{y_1} &=& - \sigma^{y_1}{}_{y_2 y_3} u^{y_2} u^{y_3} - \sigma^{y_1}{}_{y_2} a^{y_2} \nonumber \\
&&- 2 \sigma^{y_1}{}_{y_2 x_3} u^{y_2} \left( K^{x_3}{}_{y_4} u^{y_4} - H^{x_3}{}_{y_4} \frac{D \sigma^{y_4}}{dt} \right) \nonumber \\
&& - \sigma^{y_1}{}_{x_2 x_3}  \left( K^{x_2}{}_{y_4} u^{y_4} - H^{x_2}{}_{y_4} \frac{D \sigma^{y_4}}{dt} \right) \nonumber \\
&& \times \left( K^{x_3}{}_{y_5} u^{y_5} - H^{x_3}{}_{y_5} \frac{D \sigma^{y_5}}{dt} \right) \nonumber \\
&& - \sigma^{y_1}{}_{x_2} \frac{D}{dt} \left( K^{x_2}{}_{y_3} u^{y_3} - H^{x_2}{}_{y_3} \frac{D \sigma^{y_3}}{dt} \right). \label{eta_2nd_deriv_alternative_1}
\end{eqnarray}
We may derive an alternative version of this equation -- by using (\ref{formal_vel_1}) multiplied by $d t / d \tilde{t}$ -- which yields
\begin{eqnarray}
\tilde{u}^{x_3} &=& K^{x_3}{}_{y_2} u^{y_2} \frac{dt}{d\tilde{t}} - H^{x_3}{}_{y_1} \frac{D \sigma^{y_1}}{dt} \frac{dt}{d\tilde{t}}, \label{formal_vel_2}
\end{eqnarray} 
and inserted into (\ref{eta_2nd_deriv}):
\begin{eqnarray}
\frac{D^2}{dt^2} \eta^{y_1} &=& - \sigma^{y_1}{}_{y_2 y_3} u^{y_2} u^{y_3} - \sigma^{y_1}{}_{y_2} a^{y_2} - \sigma^{y_1}{}_{x_2} \tilde{a}^{x_2} \left(\frac{d\tilde{t}}{dt} \right)^2  \nonumber \\
&&- 2 \sigma^{y_1}{}_{y_2 x_3} u^{y_2} \left( K^{x_3}{}_{y_4} u^{y_4} - H^{x_3}{}_{y_4} \frac{D \sigma^{y_4}}{dt} \right) \nonumber \\
&& - \sigma^{y_1}{}_{x_2 x_3}  \left( K^{x_2}{}_{y_4} u^{y_4} - H^{x_2}{}_{y_4} \frac{D \sigma^{y_4}}{dt} \right) \nonumber \\
&& \times \left( K^{x_3}{}_{y_5} u^{y_5} - H^{x_3}{}_{y_5} \frac{D \sigma^{y_5}}{dt} \right) \nonumber \\
&& - \sigma^{y_1}{}_{x_2} \frac{d t}{d \tilde{t}} \frac{d^2\tilde{t}}{dt^2} \left( K^{x_2}{}_{y_3} u^{y_3} - H^{x_2}{}_{y_3} \frac{D \sigma^{y_3}}{dt} \right). \label{eta_2nd_deriv_alternative_2}
\end{eqnarray}
Note that we may determine the factor $d \tilde{t} / dt $ by requiring that the velocity along the curve $X$ is normalized, i.e.\ $\tilde{u}^x \tilde{u}_x =1$, in which case (\ref{formal_vel_1}) yields
\begin{eqnarray}
\frac{d\tilde{t}}{dt} &=& \tilde{u}_{x_1} K^{x_1}{}_{y_2} u^{y_2} -  \tilde{u}_{x_1} H^{x_1}{}_{y_2} \frac{D \sigma^{y_2}}{dt}. \label{formal_vel_3}
\end{eqnarray} 

\subsection{Expansion of quantities on $Y$}

The generalized (exact) deviation equations (\ref{eta_2nd_deriv_alternative_1}) and (\ref{eta_2nd_deriv_alternative_2}) contain quantities which are not defined along the reference curve, in particular the covariant derivatives of the world function. We now make use of the covariant expansions of these quantities, which we already worked out in our previous paper \cite{Puetzfeld:Obukhov:2014:2}, i.e.\
\begin{eqnarray}
\sigma^{y_0}{}_{x_1} &=& g^{y'}{}_{x_1}\biggl( -\,\delta^{y_0}{}_{y'}\nonumber\\
&& +\,\sum\limits_{k=2}^\infty\,{\frac {1}{k!}}\,\alpha^{y_0}{}_{y'y_2\!\dots \!y_{k+1}}\sigma^{y_2}\cdots\sigma^{y_{k+1}}\biggr)\!,\label{app_expansion_1}\\
\sigma^{y_0}{}_{y_1} &=& \delta^{y_0}{}_{y_1} \nonumber\\
&& -\,\sum\limits_{k=2}^\infty\,{\frac {1}{k!}}\,\beta^{y_0}{}_{y_1y_2\dots y_{k+1}} \sigma^{y_2}\!\cdots\!\sigma^{y_{k+1}}, \label{app_expansion_2} \\
g^{y_0}{}_{x_1 ; x_2} &=& g^{y'}{\!}_{x_1} g^{y''}{\!}_{x_2}\biggl({\frac 12} 
R^{y_0}{}_{y'y''y_3}\sigma^{y_3}\nonumber\\ 
&&\!+\!\sum\limits_{k=2}^\infty\,{\frac {1}{k!}}\,\gamma^{y_0}{}_{y'y''y_3\dots y_{k+2}}\sigma^{y_3}\!\cdots\!\sigma^{y_{k+2}}\!\biggr)\!,\label{app_expansion_3} \\
g^{y_0}{}_{x_1 ; y_2} &=& g^{y'}{\!}_{x_1} \biggl({\frac 12} R^{y_0}{}_{y'y_2y_3}\sigma^{y_3}\nonumber\\ 
&&\!+\!\sum\limits_{k=2}^\infty\,{\frac {1}{k!}}\,\gamma^{y_0}{}_{y'y_2y_3\dots y_{k+2}}\sigma^{y_3}\!\cdots\!\sigma^{y_{k+2}}\!\biggr).\label{app_expansion_4}
\end{eqnarray}
The coefficients $\alpha, \beta, \gamma$ in these expansions are polynomials constructed from the Riemann curvature tensor and its covariant derivatives. The first coefficients read (as one can also check using computer algebra \cite{Ottewill:2011}):
\begin{eqnarray}
\alpha^{y_0}{}_{y_1y_2y_3} &=& -\,\frac{1}{3} R^{y_0}{}_{(y_2y_3)y_1},\label{a1}\\
\beta^{y_0}{}_{y_1y_2y_3} &=& \frac{2}{3}R^{y_0}{}_{(y_2y_3)y_1},\label{be1}\\
\alpha^{y_0}{}_{y_1y_2y_3y_4} &=& \frac{1}{2} \nabla_{(y_2}R^{y_0}{}_{y_3y_4)y_1},\label{al2}\\
\beta^{y_0}{}_{y_1y_2y_3y_4} &=& -\,\frac{1}{2} \nabla_{(y_2} R^{y_0}{}_{y_3y_4)y_1},\label{be2}\\
\alpha^{y_0}{}_{y_1y_2y_3y_4y_5}  &=& -\,\frac{7}{15} R^{y_0}{}_{(y_2y_3|y'|} R^{y'}{}_{y_4y_5)y_1}  \nonumber \\
&& -\,\frac{3}{5} \nabla_{(y_5} \nabla_{y_4} R^{y_0}{}_{y_2y_3)y_1},\label{al3}  \\
\beta^{y_0}{}_{y_1y_2y_3y_4y_5} &=& \frac{8}{15} R^{y_0}{}_{(y_2y_3|y'|} R^{y'}{}_{y_4y_5)y_1} \nonumber \\
&&+\,\frac{2}{5}  \nabla_{(y_5} \nabla_{y_4} R^{y_0}{}_{y_2y_3)y_1} ,\label{be3}\\
\gamma^{y_0}{}_{y_1y_2y_3y_4}&=& \frac{1}{3} \nabla_{(y_3} R^{y_0}{}_{|y_1|y_4)y_2}, \label{ga} \\
\gamma^{y_0}{}_{y_1y_2y_3y_4y_5} &=& \frac{1}{4}  R^{y_0}{}_{y_1y'(y_3} R^{y'}{}_{y_4y_5)y_2} \nonumber \\
&& +\,\frac{1}{4} \nabla_{(y_5}\nabla_{y_4} R^{y_0}{}_{|y_1y_2|y_3)}. \label{ga2}
\end{eqnarray}
These results allow us to derive the third derivatives of the world function appearing in (\ref{eta_2nd_deriv_alternative_1}) and (\ref{eta_2nd_deriv_alternative_2}), i.e.\ we have up to the second order in the deviation vector:
\begin{eqnarray}
\sigma^{y_0}{}_{y_1 y_2} &=& - \frac{2}{3} R^{y_0}{}_{(y_2 y_3) y_1} \sigma^{y_3} - \frac{1}{2} \left( \frac{1}{2}  \nabla_{y_2} R^{y_0}{}_{(y_3 y_4) y_1} \right. \nonumber \\
&& \left. - \frac{1}{3} \nabla_{y_3} R^{y_0}{}_{(y_2 y_4) y_1} \right)   \sigma^{y_3} \sigma^{y_4}\nonumber\\ 
&& -\,{\frac 16}\lambda^{y_0}{}_{y_1 y_2 y_3y_4y_5}\sigma^{y_3}\sigma^{y_4}\sigma^{y_5}  + {\mathcal O}(\sigma^4), \\
\sigma^{y_0}{}_{y_1 x_2} &=& g^{y'}{}_{x_2} \left( \frac{2}{3}  R^{y_0}{}_{(y' y_3) y_1} \sigma^{y_3} \right. \nonumber \\
&& -\,\frac{1}{4} \nabla_{(y'} R^{y_0}{}_{y_3 y_4) y_1} \sigma^{y_3} \sigma^{y_4} \nonumber\\ 
&& \left. + \,{\frac 16}\mu^{y_0}{}_{y_1 y' y_3y_4y_5}\sigma^{y_3}\sigma^{y_4}\sigma^{y_5} \right) + {\mathcal O}(\sigma^4), \\
\sigma^{y_0}{}_{x_1 x_2} &=& - g^{y'}{}_{x_1} g^{y''}{}_{x_2} \left[\left(\frac{1}{2} R^{y_0}{}_{y' y'' y_3} - \frac{1}{3} R^{y_0}{}_{( y'' y_3) y'} \right) \sigma^{y_3} \right. \nonumber \\
&& +\left( \frac{1}{6}  \nabla_{(y_3} R^{y_0}{}_{| y' | y_4 ) y''}  + \frac{1}{4} \nabla_{(y''} R^{y_0}{}_{y_3 y_4) y'} \right) \sigma^{y_3} \sigma^{y_4} \nonumber\\
&& \left. +\,{\frac 16}\nu^{y_0}{}_{y' y'' y_3y_4y_5}\sigma^{y_3}\sigma^{y_4}\sigma^{y_5} \right]  + {\mathcal O}(\sigma^4). 
\end{eqnarray}
Here we introduced a compact notation for the combinations of the second covariant derivatives of the curvature and the quadratic polynomial of the curvature tensor (in symbolic form, ``$\nabla\nabla R + R\cdot R$''):
\begin{eqnarray}
\lambda^{y_0}{}_{y_1 y_2 y_3y_4y_5} &=& \beta^{y_0}{}_{y_1 y_3 y_4y_5;y_2} + \beta^{y_0}{}_{y_1 y_2 y_3y_4y_5}\nonumber\\
&& -\,3\beta^{y_0}{}_{y_1 y' (y_3}\beta^{y'}{}_{|y_2|y_4y_5)},\label{lambda}\\
\mu^{y_0}{}_{y_1 y_2 y_3y_4y_5} &=& \beta^{y_0}{}_{y_1 y_2 y_3y_4y_5}\nonumber\\ 
&& -\,3\beta^{y_0}{}_{y_1 y' (y_3}\alpha^{y'}{}_{|y_2|y_4y_5)},\label{mu}\\
\nu^{y_0}{}_{y_1 y_2 y_3y_4y_5} &=& \gamma^{y_0}{}_{y_1 y_2y_3 y_4y_5} + \alpha^{y_0}{}_{y_1 y_2 y_3y_4y_5}\nonumber\\
&& -\,3\alpha^{y_0}{}_{y_1 y' (y_3}\alpha^{y'}{}_{|y_2|y_4y_5)}\nonumber  \\
&& -\,\frac{1}{4} R^{y'}{}_{y_1 y_2 (y_3}\alpha^{y_0}{}_{|y'|y_4 y_5)}.\label{nu}
\end{eqnarray}
Substituting the coefficients of the expansions (\ref{app_expansion_1})-(\ref{app_expansion_3}) we obtain the explicit (complicated) expressions which we do not display here.

For the symmetrized versions we obtain 
\begin{eqnarray}
\sigma^{y_0}{}_{(y_1 y_2)} &=& \frac{1}{3} R^{y_0}{}_{(y_1 y_2) y_3} \sigma^{y_3} - \frac{1}{4} \left( \nabla_{(y_1} R^{y_0}{}_{|y_3 y_4 | y_2)} \right. \nonumber \\
&& \left. + \frac{1}{3} \nabla_{y_3} R^{y_0}{}_{(y_1 y_2) y_4} \right) \sigma^{y_3} \sigma^{y_4}\nonumber\\ 
&& -\,{\frac 16}\lambda^{y_0}{}_{(y_1 y_2) y_3y_4y_5}\sigma^{y_3}\sigma^{y_4}\sigma^{y_5} + {\mathcal O}(\sigma^4),\nonumber \\
&& \\
\sigma^{y_0}{}_{(x_1 x_2)} &=& g^{y'}{}_{(x_1} g^{y''}{}_{x_2)} \left[-\frac{2}{3} R^{y_0}{}_{(y' y'') y_3} \sigma^{y_3} \right. \nonumber \\
&& \left.  + \frac{1}{4} \left(  \nabla_{y_3} R^{y_0}{}_{( y' y'' ) y_4 } \right. \right. \nonumber \\
&&\left. -\,\frac{1}{3} \nabla_{(y'} R^{y_0}{}_{|y_3 y_4 | y'')} \right) \sigma^{y_3} \sigma^{y_4} \nonumber \\ 
&&\left. -\,{\frac 16}\nu^{y_0}{}_{(y' y'') y_3y_4y_5}\sigma^{y_3}\sigma^{y_4}\sigma^{y_5}\right] + {\mathcal O}(\sigma^4),\nonumber \\
\end{eqnarray}
Furthermore we need the expansions of $K^{x}{}_{y}$ and $H^{x}{}_{y}$. We already have everything at hand except $\stackrel{-1}{\sigma}{}\!\!^{x}{}_y$ which we can obtain from (\ref{inverse_2}):  
\begin{eqnarray}
\stackrel{-1}{\sigma}{}\!\!^{x_1}{}_{y_2} &=& - H^{x_1}{}_{y_2} \nonumber \\
&=& - g^{x_1}{}_{y'} \left( \delta^{y'}{}_{y_2} + \sum\limits_{k=2}^\infty\,{\frac {1}{k!}} h^{y'}{}_{y_2y_3\dots y_{k+2}} \sigma^{y_2}\!\cdots\!\sigma^{y_{k+2}} \right) \nonumber\\
&=& - g^{x_1}{}_{y'} \left( \delta^{y'}{}_{y_2} - \frac{1}{6} R^{y'}{}_{(y_3 y_4) y_2} \sigma^{y_3} \sigma^{y_4}  \right. \nonumber \\
&& \left. + \frac{1}{12} \nabla_{(y_3} R^{y'}{}_{y_4 y_5)y_2}  \sigma^{y_3} \sigma^{y_4} \sigma^{y_5} \right) + {\mathcal O}(\sigma^4) ,\\
 K^{x_1}{}_{y_2}&=& g^{x_1}{}_{y'} \left( \delta^{y'}{}_{y_2} + \sum\limits_{k=2}^\infty\,{\frac {1}{k!}} k^{y'}{}_{y_2y_3\dots y_{k+2}} \sigma^{y_2}\!\cdots\!\sigma^{y_{k+2}} \right) \nonumber\\
&=& g^{x_1}{}_{y'} \left( \delta^{y'}{}_{y_2} - \frac{1}{2} R^{y'}{}_{(y_3 y_4) y_2} \sigma^{y_3} \sigma^{y_4} \right. \nonumber \\
&& \left. +\,\frac{1}{6} \nabla_{(y_3} R^{y'}{}_{y_4 y_5)y_2}  \sigma^{y_3} \sigma^{y_4} \sigma^{y_5} \right) + {\mathcal O}(\sigma^4).
\end{eqnarray}
From this one can derive the recurring term in (\ref{eta_2nd_deriv_alternative_2}) up to the needed order, i.e.
\begin{eqnarray}
&& \left( K^{x_1}{}_{y_2} u^{y_2} - H^{x_1}{}_{y_2} \frac{D \sigma^{y_2}}{dt} \right) 
=  g^{x_1}{}_{y'} \Biggl[ u^{y'} - \frac{D \sigma^{y'}}{dt}\nonumber \\
&& - \,\frac{1}{2} R^{y'}{}_{(y_3 y_4) y_2} \sigma^{y_3} \sigma^{y_4} \left( u^{y_2} - \frac{1}{3}\frac{D \sigma^{y_2}}{dt}\right) \nonumber\\
&& +\,\frac{1}{6}\nabla_{(y_3} R^{y'}{}_{y_4 y_5)y_2} u^{y_2}\sigma^{y_3} \sigma^{y_4} \sigma^{y_5}\Biggr]  + {\mathcal O}(\sigma^4).
\end{eqnarray}
With these expansions at hand we are finally able to develop the deviation equation (\ref{eta_2nd_deriv_alternative_2}) up to the third order.

Denote $\tilde{a}^{y_1} = g^{y_1}{}_{x_2} \tilde{a}^{x_2}$ in accordance with the definition of the parallel propagator, and introduce 
\begin{eqnarray}
\phi^{y_1}{}_{y_2 y_3 y_4y_5y_6} &=& \lambda^{y_1}{}_{y_2 y_3 y_4y_5y_6} - 2\mu^{y_1}{}_{y_2 y_3 y_4y_5y_6} \nonumber \\  && + \nu^{y_1}{}_{y_2 y_3 y_4y_5y_6}. \label{phi}
\end{eqnarray}
\begin{widetext}
The deviation equation up to the third order reads
\begin{eqnarray}
&& \frac{D^2}{dt^2} \eta^{y_1} = \tilde{a}^{y_1} \left(\frac{d\tilde{t}}{dt} \right)^2 - a^{y_1} +  \frac{d t}{d \tilde{t}} \frac{d^2\tilde{t}}{dt^2} u^{y_1} + \frac{D \eta^{y_1}}{dt} \frac{d t}{d \tilde{t}} \frac{d^2\tilde{t}}{dt^2}
 - \eta^{y_4} R^{y_1}{}_{y_2 y_3 y_4} \left(  u^{y_2} u^{y_3} + 2 u^{y_3} \frac{D \eta^{y_2}}{dt} \right) \nonumber \\
&&  + \eta^{y_4} \eta^{y_5} \left\{ u^{y_2} u^{y_3}  \left( \frac{1}{2} \nabla_{y_2} R^{y_1}{}_{y_4 y_5 y_3 } - \frac{1}{3}  \nabla_{y_4} R^{y_1}{}_{y_2 y_3 y_5} \right) +  \frac{1}{3} R^{y_1}{}_{y_4 y_5 y_2} \left[ a^{y_2} + \frac{1}{2} \tilde{a}^{y_2} \left(\frac{d\tilde{t}}{dt} \right)^2 - u^{y_2} \frac{d t}{d \tilde{t}} \frac{d^2\tilde{t}}{dt^2} \right]\right\} \nonumber \\
&&  - {\frac 16}\eta^{y_4} \eta^{y_5} \eta^{y_6} \left\{ \phi^{y_1}{}_{y_2 y_3 y_4y_5y_6} u^{y_2}u^{y_3} - {\frac 12}\nabla_{(y_4}R^{y_1}{}_{y_5y_6)y_2} \left[ a^{y_2} + \tilde{a}^{y_2} \left(\frac{d\tilde{t}}{dt} \right)^2 - u^{y_2} \frac{d t}{d \tilde{t}} \frac{d^2\tilde{t}}{dt^2} \right]\right\}\nonumber \\
&& - {\frac 12}u^{y'}\frac{D \eta^{y''}}{dt}\eta^{y_2}\eta^{y_3}\left( - \nabla_{(y''} R^{y_1}{}_{y_2 y_3) y_1 } + \nabla_{y_2} R^{y_1}{}_{(y' y'') y_3} - {\frac 13}\nabla_{(y'} R^{y_1}{}_{|y_2 y_3|y'')} \right) - {\frac 23}\frac{D \eta^{y_2}}{dt}\frac{D \eta^{y_3}}{dt}\eta^{y_4}\,R^{y_1}{}_{y_2y_3y_4} + \mathcal{O}(\sigma^4).\nonumber\\ \label{eta_2nd_deriv_compact_2}
\end{eqnarray}
\end{widetext}
We would like to stress that the generalized deviation equation derived in (\ref{eta_2nd_deriv_compact_2}) is completely general. In particular, it allows for a comparison of two general, i.e.\ not necessarily geodetic, world lines in spacetime. 

In special cases, cf.\ also the next section \ref{sec_special_dev_derivation}, our result is in qualitative agreement with previous results in the literature, see in particular \cite{Hodgkinson:1972,Bazanski:1977:1,Aleksandrov:Piragas:1978,Schutz:1985,Mullari:Tammelo:2006,Vines:2014}

\section{Special cases}\label{sec_special_dev_derivation}

Up to this point our considerations were completely general, resulting in the exact form (\ref{eta_2nd_deriv_alternative_2}) as well as in the second order version (\ref{eta_2nd_deriv_compact_2}) -- expanded w.r.t. the world function -- of the generalized deviation equation. In the following we will study some special cases of the deviation equation.

\subsection{Affine parametrization}

So-far our framework allows for a completely general parametrization of the curves $Y$ and $X$. While such a general framework is of course desirable from a mathematical point of view, such freedom of the parametrization may also lead to unnecessarily complicated equations. By switching to an affine parametrization of the curves, i.e.\ we assume that the time parameter $\tilde{t}$ on $X$ is a linear function of the one on $Y$, we can simplify the deviation equation, without restricting its physical meaning. If we demand that $\tilde{t}=c_1 t + c_2$, where $c_1$ and $c_2$ are just some arbitrary constants, we can get rid of the ``parametrization induced'' acceleration terms.  In particular, the exact deviation equation (\ref{eta_2nd_deriv_alternative_2}) now takes the form:
\begin{eqnarray}
\frac{D^2}{dt^2} \eta^{y_1} &=& - \sigma^{y_1}{}_{y_2 y_3} u^{y_2} u^{y_3} - \sigma^{y_1}{}_{y_2} a^{y_2}  - \sigma^{y_1}{}_{x_2} \ap^{x_2} \nonumber \\
&&- 2 \sigma^{y_1}{}_{y_2 x_3} u^{y_2} \left( K^{x_3}{}_{y_4} u^{y_4} - H^{x_3}{}_{y_4} \frac{D \sigma^{y_4}}{dt} \right) \nonumber \\
&& - \sigma^{y_1}{}_{x_2 x_3}  \left( K^{x_2}{}_{y_4} u^{y_4} - H^{x_2}{}_{y_4} \frac{D \sigma^{y_4}}{dt} \right) \nonumber \\
&& \times \left( K^{x_3}{}_{y_5} u^{y_5} - H^{x_3}{}_{y_5} \frac{D \sigma^{y_5}}{dt} \right) . \label{eta_2nd_deriv_alternative_2_affine}
\end{eqnarray}
Note that here we introduced the new symbol $\ap^{x}$ for the acceleration on $X$, to distinguish it from the acceleration $\widetilde{a}^x$ for an arbitrary parametrization in the original equation  (\ref{eta_2nd_deriv_alternative_2}). Furthermore, for an affine parametrization the approximated version (\ref{eta_2nd_deriv_compact_2}) of the generalized deviation takes the following simplified form:
\begin{eqnarray}
&& \frac{D^2}{dt^2} \eta^{y_1} = \ap^{y_1} - a^{y_1}  \nonumber \\
&&- \eta^{y_4} R^{y_1}{}_{y_2 y_3 y_4} \left(  u^{y_2} u^{y_3} + 2 u^{y_3} \frac{D \eta^{y_2}}{dt} \right) \nonumber \\
&&  + \eta^{y_4} \eta^{y_5} \left\{ u^{y_2} u^{y_3}  \left( \frac{1}{2} \nabla_{y_2} R^{y_1}{}_{y_4 y_5 y_3 } - \frac{1}{3}  \nabla_{y_4} R^{y_1}{}_{y_2 y_3 y_5} \right) \right. \nonumber \\
&& \left. +  \frac{1}{3} R^{y_1}{}_{y_4 y_5 y_2} \left[ a^{y_2} + \frac{1}{2} \ap^{y_2} \right]\right\} + \mathcal{O}(\sigma^3). \label{eta_2nd_deriv_compact_2_affine}
\end{eqnarray}

\subsubsection{Geodesic curves}

If the two curves $Y$ and $X$ are geodesics, then (\ref{eta_2nd_deriv_compact_2_affine}) takes the even simpler form:
\begin{eqnarray}
&& \frac{D^2}{dt^2} \eta^{y_1} = - \eta^{y_4} R^{y_1}{}_{y_2 y_3 y_4} \left(  u^{y_2} u^{y_3} + 2 u^{y_3} \frac{D \eta^{y_2}}{dt} \right) \nonumber \\
&&  + \eta^{y_4} \eta^{y_5} \left\{ u^{y_2} u^{y_3}  \left( \frac{1}{2} \nabla_{y_2} R^{y_1}{}_{y_4 y_5 y_3 } - \frac{1}{3}  \nabla_{y_4} R^{y_1}{}_{y_2 y_3 y_5} \right) \right\}\nonumber \\ 
&&+ \mathcal{O}(\sigma^3). \label{eta_2nd_deriv_compact_2_affine_geodesics}
\end{eqnarray}

Furthermore, from (\ref{eta_2nd_deriv_compact_2_affine_geodesics}) we can recover the well-known equation of geodesic deviation by linearizing in $\eta$:  
\begin{eqnarray}
\frac{D^2}{dt^2} \eta^{y_1} &=& - R^{y_1}{}_{y_2 y_3 y_4} u^{y_2} u^{y_3} \eta^{y_4}. \label{geodesic_dev_revovered_affine}
\end{eqnarray}

\subsubsection{Flat spacetime}

In a flat spacetime, and for affine parametrization, equation (\ref{eta_2nd_deriv_compact_2_affine}) yields:
\begin{eqnarray}
\frac{D^2}{dt^2} \eta^{y} = \ap^{y} - a^{y}. 
\end{eqnarray}
Hence, if the two curves are geodesics, we obtain the expected result
\begin{eqnarray}
\frac{D^2}{dt^2} \eta^{y} = 0. 
\end{eqnarray}

\subsection{Synchronous parametrization}

The factors with the derivatives of the parameters $t$ and $\tilde{t}$ appear due to the non-synchronous parametrization of the two curves. It is possible to make things simpler by introducing the synchronization of parametrization. Namely, we start by rewriting the velocity as
\begin{equation}
u^y = {\frac {dY^y}{dt}} = {\frac {d\tilde{t}}{dt}} {\frac {dY^y}{d\tilde{t}}}.\label{ut1}
\end{equation}
That is, we now parametrize the position on the first curve by the same variable $\tilde{t}$ that is used on the second curve. Accordingly, we denote 
\begin{equation}
\ut^y = {\frac {dY^y}{d\tilde{t}}}.\label{ut2}
\end{equation} 
By differentiation, we then derive
\begin{eqnarray}
a^y &=& {\frac {Du^y}{dt}} = {\frac {D}{dt}}\left( {\frac {d\tilde{t}}{dt}}\ut^y\right)\nonumber\\
&=& {\frac {d^2\tilde{t}}{dt^2}}\ut^y + \left({\frac {d\tilde{t}}{dt}}\right)^2\at^y,\label{at1}
\end{eqnarray}
where 
\begin{equation}
\at^y = {\frac {D}{d\tilde{t}}}\ut^y = {\frac {D^2Y^y}{d\tilde{t}^2}}.\label{at2}
\end{equation}
Analogously, we derive for the derivative of the deviation vector
\begin{equation}
{\frac {D^2\eta^y}{dt^2}} =  {\frac {d^2\tilde{t}}{dt^2}}{\frac {D\eta^y}{d\tilde{t}}}
+ \left({\frac {d\tilde{t}}{dt}}\right)^2 {\frac {D^2\eta^y}{d\tilde{t}^2}}.\label{Deta2}
\end{equation}
Substituting (\ref{at1}) and (\ref{Deta2}) into (\ref{eta_2nd_deriv_compact_2}), we obtain 
\begin{eqnarray}
&& \frac{D^2}{d\tilde{t}^2} \eta^{y_1} = \tilde{a}^{y_1} - \at^{y_1}  
 \nonumber \\
&& - \eta^{y_4} R^{y_1}{}_{y_2 y_3 y_4} \left( \ut^{y_2} \ut^{y_3} + 2 \ut^{y_3} \frac{D \eta^{y_2}}{d\tilde{t}} \right) \nonumber \\
&&  + \eta^{y_4} \eta^{y_5} \left\{ \ut^{y_2} \ut^{y_3}  \left( \frac{1}{2} \nabla_{y_2} R^{y_1}{}_{y_4 y_5 y_3 } - \frac{1}{3}  \nabla_{y_4} R^{y_1}{}_{y_2 y_3 y_5} \right) \right. \nonumber \\
&& \left. +  \frac{1}{3} R^{y_1}{}_{y_4 y_5 y_2} \left( \at^{y_2} + \frac{1}{2} \tilde{a}^{y_2} \right)\right\} + \mathcal{O}(\sigma^3). \label{eta_2nd_deriv_compact_3}
\end{eqnarray}

Now everything is synchronous in the sense that both curves are parametrized by $\tilde{t}$.

Actually, the synchronization can be done already for the exact deviation equation (\ref{eta_2nd_deriv_alternative_2}) which is then recast into a simpler form
\begin{eqnarray}
\frac{D^2}{d\tilde{t}^2} \eta^{y_1} &=& - \sigma^{y_1}{}_{y_2} \at^{y_2} - \sigma^{y_1}{}_{x_2} \tilde{a}^{x_2} - \sigma^{y_1}{}_{y_2 y_3} \ut^{y_2} \ut^{y_3} \nonumber \\
&&- 2 \sigma^{y_1}{}_{y_2 x_3} \ut^{y_2} \left( K^{x_3}{}_{y_4} \ut^{y_4} - H^{x_3}{}_{y_4} \frac{D \sigma^{y_4}}{d\tilde{t}} \right) \nonumber \\
&& - \sigma^{y_1}{}_{x_2 x_3}  \left( K^{x_2}{}_{y_4} \ut^{y_4} - H^{x_2}{}_{y_4} \frac{D \sigma^{y_4}}{d\tilde{t}} \right) \nonumber \\
&& \times \left( K^{x_3}{}_{y_5} \ut^{y_5} - H^{x_3}{}_{y_5} \frac{D \sigma^{y_5}}{d\tilde{t}} \right). \label{eta_2nd_deriv_alternative_3}
\end{eqnarray}

\subsubsection{Geodesic curves}

If the two curves $Y$ and $X$ are geodesics, then (\ref{eta_2nd_deriv_compact_2}) takes the form:
\begin{eqnarray}
&&\frac{D^2}{d\tilde{t}^2} \eta^{y_1} =  - \eta^{y_4} R^{y_1}{}_{y_2 y_3 y_4} \left(  \ut^{y_2} \ut^{y_3} + 2 \ut^{y_3} \frac{D \eta^{y_2}}{d\tilde{t}} \right) \nonumber \\
&&+\,\eta^{y_4} \eta^{y_5} \ut^{y_2} \ut^{y_3}  \left( \frac{1}{2} \nabla_{y_2} R^{y_1}{}_{y_4 y_5 y_3 } - \frac{1}{3}  \nabla_{y_4} R^{y_1}{}_{y_2 y_3 y_5} \right). \nonumber \\ \label{gendev_syn_geo}
\end{eqnarray}
This equation allows for a direct comparison to several previous results in the literature. In particular it is in qualitative agreement, note the difference in some prefactors, with \cite[(2.51)]{Hodgkinson:1972}, \cite[(4.2)]{Bazanski:1977:1}, \cite[(D1,D2)]{Aleksandrov:Piragas:1978}, \cite[(39)]{Schutz:1985}. 

\subsubsection{Flat spacetime}

In a flat spacetime, in Cartesian coordinates, the deviation equation (\ref{eta_2nd_deriv_compact_2}) takes the form:
\begin{eqnarray}
\frac{D^2}{d\tilde{t}^2} \eta^{y_1} = \tilde{a}^{y_1}. 
\end{eqnarray}
This can be recast into 
\begin{eqnarray}
\frac{D^2}{d\tilde{t}^2}\left( \eta^{y_1} - Y^{y_1} + X^{y_1}\right) = 0.
\end{eqnarray}
Taking into account the definitions and the initial conditions, we conclude that 
\begin{equation}
\eta^{y_1} = Y^{y_1} - X^{y_1},
\end{equation}
which is what we expect -- we are thus recovering the definition (\ref{gen_dev_definition}). 

\subsection{Orthogonal parametrization}

Its is worthwhile to stress that no assumption about the orthogonality of the deviation vector $\eta^y$ w.r.t.\ the velocity $u^y$ along the reference world line has been made in our derivation. Such an additional assumption could be imposed, basically leading to a form of the deviation equation as given in \cite{Bazanski:1977:1}. Technically, this is achieved by performing an orthogonal decomposition of the generalized deviation equation. This is straightforward and we do not present here the explicit result.

\subsection{Flat spacetime, geodesic curves}

In flat spacetime, and for the curves $Y$ and $X$ being geodesics, we obtain:
\begin{eqnarray}
\frac{D^2}{dt^2} \eta^{y_1} = \frac{d t}{d \tilde{t}} \frac{d^2\tilde{t}}{dt^2} u^{y_1} + \frac{D \eta^{y_1}}{dt} \frac{d t}{d \tilde{t}} \frac{d^2\tilde{t}}{dt^2}. 
\end{eqnarray}
In order to arrive at the intuitive result of a non accelerated deviation vector, we have to make sure there is no ``parametrization induced'' acceleration, once again by choosing the parametrization in such a way that ${d^2\tilde{t}}/{dt^2}$ vanishes. In the synchronized form, we have
\begin{eqnarray}
\frac{D^2}{d\tilde{t}^2} \eta^{y_1} = 0.
\end{eqnarray}

\section{Gravitational compass}\label{sec_compass}

The determination of the curvature of spacetime in the context of deviation equations has been discussed in previous works, see for example \cite{Synge:1960,Szekeres:1965,Ciufolini:Demianski:1986}. In \cite{Szekeres:1965}, Szekeres coined the notion of a ``gravitational compass.'' From now on we will adopt this notion for a set of suitably prepared test bodies which allow for the measurement of the curvature and, thereby, the gravitational field. 

The operational procedure is to monitor the accelerations of a set of test bodies w.r.t.\ to an observer moving on the reference world line $Y$. A mechanical analogue would be to measure the forces between the test bodies and the reference body via a spring connecting them. 

In the following we search for configurations of test bodies which allow for a complete determination of all curvature components in a Riemannian background spacetime. We perform our analysis on the basis of the standard geodesic deviation equation, as well as one of its generalizations.

\subsection{Rewriting the deviation equation}\label{sec_compass_with_standard_dev}

Our starting point is the standard geodesic deviation equation, i.e. 
\begin{eqnarray}
\frac{D^2}{ds^2}\eta^a = R^a{}_{bcd} u^b \eta^c u^d. \label{compass_start}
\end{eqnarray}
Since we want to express the curvature in terms of measured quantities, i.e.\ the velocities and the accelerations, we rewrite this equation in terms of the standard (non-covariant) derivative w.r.t.\ the proper time.

In order to simplify the resulting equation we employ normal coordinates, i.e.\ we have on the world line of the reference test body
\begin{eqnarray}
\Gamma_{ab}{}^c |_Y= 0, \quad \quad \partial_a \Gamma_{bc}{}^d|_Y = \frac{2}{3} R_{a(bc)}{}^d.
\end{eqnarray}

In terms of the standard total derivative w.r.t.\ to the proper time $s$, the deviation equation (\ref{compass_start}) takes the form:
\begin{eqnarray}
\frac{d^2}{ds^2}\eta^a &\stackrel{|_Y}{=}& \frac{2}{3} R^a{}_{bcd} u^b \eta^c u^d. \label{compass_upper}
\end{eqnarray}
However, what actually seems to be measured by a compass at the reference point $Y$ is the lower components of the relative acceleration. For the lower index position, in terms of the ordinary derivative in normal coordinates, the deviation equation (\ref{compass_start}) takes the form
\begin{eqnarray}
\frac{d^2}{ds^2}\eta_a &\stackrel{|_Y}{=}& \frac{4}{3} R_{abcd} u^b \eta^c u^d. \label{compass_lower}
\end{eqnarray} 

\subsection{Explicit compass setup}\label{sec_compass_setup_standard_dev}

Let us consider a general 6-point compass. In addition to the reference test body on the world line we will use the following geometrical setup of the 5 remaining test bodies:
\begin{eqnarray}
&&{}^{(1)}\eta^{a}=\left(\begin{array}{c} 0 \\ 1\\ 0\\ 0\\ \end{array} \right),
{}^{(2)}\eta^{a}=\left(\begin{array}{c} 0 \\ 0\\ 1\\ 0\\ \end{array} \right),
{}^{(3)}\eta^{a}=\left(\begin{array}{c} 0 \\ 0\\ 0\\ 1\\ \end{array} \right), \nonumber \\
&&{}^{(4)}\eta^{a}=\left(\begin{array}{c} 0 \\ 1\\ 1\\ 0\\ \end{array} \right),
{}^{(5)}\eta^{a}=\left(\begin{array}{c} 0 \\ 0\\ 1\\ 1\\ \end{array} \right). \label{position_setup}
\end{eqnarray}
In addition to the positions of the compass constituents, we have to make a choice for the velocity of the reference test body / observer. In the following we will use $(m)$ different compasses, each of these compasses will have a different velocity (associated) with the reference test body. In other words, we consider $(m)$ different compasses or reference test bodies, all of which are located at the world line reference point $Y$ (at the same time), and all these $(m)$ observers measure the relative accelerations to all five test bodies placed at the positions given in (\ref{position_setup}). The lhs of (\ref{compass_lower}) are the measured accelerations and in the following we refer to them by ${}^{(m,n)}A_a$. Furthermore, we also introduced the compass index ${}^{(m)}u^a$ for the velocities. In other words, for $(m)$ compasses and $(n)$ bodies in one compass, we have the following set of equations:
\begin{eqnarray}
{}^{(m,n)}A_a &\stackrel{|_Y}{=}& \frac{4}{3} R_{abcd} {}^{(m)}u^b \, {}^{(n)}\eta^c \, {}^{(m)}u^d. \label{compass_lower_setup}
\end{eqnarray}  
What remains to be chosen, apart from the $(n=1 \dots 5)$ positions of bodies in one compass, is the number $(m)$ and the actual directions in which each compass / observer shall move. Of course in the end we want to minimize both numbers, i.e.\ $(m)$ and $(n)$, which are needed to determine all curvature components.  
\begin{eqnarray}
&&{}^{(1)}u^{a}=\left(\begin{array}{c} c_{10} \\ 0 \\ 0\\ 0\\ \end{array} \right),
{}^{(2)}u^{a}=\left(\begin{array}{c}  c_{20} \\ c_{21} \\ 0 \\ 0\\ \end{array} \right),
{}^{(3)}u^{a}=\left(\begin{array}{c}  c_{30} \\ 0 \\ c_{32} \\ 0 \\ \end{array} \right), \nonumber \\
&&{}^{(4)}u^{a}=\left(\begin{array}{c}  c_{40} \\ 0 \\ 0 \\ c_{43}\\ \end{array} \right),
{}^{(5)}u^{a}=\left(\begin{array}{c}  c_{50} \\ c_{51} \\ c_{52} \\ 0\\ \end{array} \right),
{}^{(6)}u^{a}=\left(\begin{array}{c}  c_{60} \\ 0 \\ c_{62} \\ c_{63} \\ \end{array} \right). \nonumber\\
\label{velocity_setup}
\end{eqnarray}
The $c_{(m)a}$ here are just constants, chosen appropriately to ensure the normalization of the 4-velocity of each compass. 

In summary, we are going to consider $(m)=1 \dots 6$ compasses, each of them with 6-points, where the five reference points are always the $(n)=1 \dots 5$ from (\ref{position_setup}).

\subsubsection{Explicit curvature components}

The 20 independent components of the curvature tensor can be explicitly determined in terms of the accelerations ${}^{(m,n)}A_a$ and velocities ${}^{(m)}u^a$ by making use of the deviation equation (\ref{compass_lower_setup}) with the help of the compass configuration given in (\ref{position_setup}) and (\ref{velocity_setup}). The result reads as follows:

\begin{eqnarray}
01 : R_{1010} &=& \frac{3}{4} {}^{(1,1)}A_1 c^{-2}_{10}, \label{exR1010} \\
02 : R_{2010} &=& \frac{3}{4} {}^{(1,1)}A_2 c^{-2}_{10}, \label{exR2010} \\
03 : R_{3010} &=& \frac{3}{4} {}^{(1,1)}A_3 c^{-2}_{10}, \label{exR3010} \\
04 : R_{2020} &=& \frac{3}{4} {}^{(1,2)}A_2 c^{-2}_{10}, \label{exR2020} \\
05 : R_{3020} &=& \frac{3}{4} {}^{(1,2)}A_3 c^{-2}_{10}, \label{exR3020} \\
06 : R_{3030} &=& \frac{3}{4} {}^{(1,3)}A_3 c^{-2}_{10}, \label{exR3030} \\
07 : R_{2110} &=& \frac{3}{4} {}^{(2,1)}A_2 c^{-1}_{21} c^{-1}_{20} -  R_{2010} c^{-1}_{21} c_{20} , \label{exR2110} \\
08 : R_{3110} &=& \frac{3}{4} {}^{(2,1)}A_3 c^{-1}_{21} c^{-1}_{20} -  R_{3010} c^{-1}_{21} c_{20} , \label{exR3110} \\
09 : R_{0212} &=& \frac{3}{4} {}^{(3,1)}A_0 c^{-2}_{32} +  R_{2010} c^{-1}_{32} c_{30} , \label{exR0212} \\
10 : R_{1212} &=& \frac{3}{4} {}^{(2,2)}A_2 c^{-2}_{21} -  R_{2020} c^{2}_{20} c^{-2}_{21}  -  2 R_{0212} c^{-1}_{21} c_{20}  , \nonumber \\ \label{exR1212} \\
11 : R_{3220} &=& \frac{3}{4} {}^{(3,2)}A_3 c^{-1}_{32} c^{-1}_{30} -  R_{3020} c^{-1}_{32} c_{30} , \label{exR3220} \\
12 : R_{0313} &=& \frac{3}{4} {}^{(4,1)}A_0 c^{-2}_{43} +  R_{3010} c^{-1}_{43} c_{40} , \label{exR0313} \\
13 : R_{1313} &=& \frac{3}{4} {}^{(2,3)}A_3 c^{-2}_{21} -  R_{3030} c^{2}_{20} c^{-2}_{21}  -  2 R_{0313} c^{-1}_{21} c_{20}  , \nonumber \\ \label{exR1313} \\
14 : R_{0323} &=& \frac{3}{4} {}^{(4,2)}A_0 c^{-2}_{43} +  R_{3020} c^{-1}_{43} c_{40} , \label{exR0323} \\
15 : R_{2323} &=& \frac{3}{4} {}^{(4,2)}A_2 c^{-2}_{43} -  R_{2020} c^{-2}_{43} c^{2}_{40}  +  2 R_{3220} c^{-1}_{43} c_{40}  , \nonumber \\ \label{exR2323} \\
16 : R_{3132} &=& \frac{3}{8} {}^{(5,3)}A_3 c^{-1}_{52} c^{-1}_{51}  -  \frac{1}{2} R_{3030} c^{-1}_{52} c^{-1}_{51} c^2_{50} \nonumber \\ 
&&  -  R_{0313} c^{-1}_{52} c_{50} -  R_{0323} c^{-1}_{51} c_{50} -  \frac{1}{2} R_{1313} c^{-1}_{52} c_{51} \nonumber \\
&&  - \frac{1}{2} R_{2323} c_{52} c^{-1}_{51} ,  \label{exR3132} \\
17 : R_{1213} &=& \frac{3}{8} {}^{(6,1)}A_1 c^{-1}_{63} c^{-1}_{62}  -  \frac{1}{2} R_{1010} c^{-1}_{63} c^{-1}_{62} c^2_{60} \nonumber \\ 
&&  +  R_{2110} c^{-1}_{63} c_{60} +  R_{3110} c^{-1}_{62} c_{60} -  \frac{1}{2} R_{1212} c^{-1}_{63} c_{62} \nonumber \\
&&  - \frac{1}{2} R_{1313} c_{63} c^{-1}_{62} ,  \label{exR1213} 
\end{eqnarray}

\begin{figure}
\begin{center}
\includegraphics[width=7cm,angle=-90]{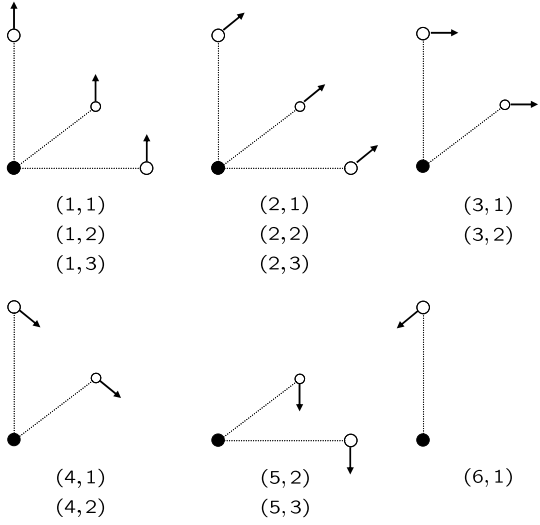}
\end{center}
\caption{\label{fig_standard_compass} Symbolical sketch of the explicit compass solution in (\ref{exR1010})-(\ref{ex3212}). In total 13 suitably prepared test bodies (hollow circles) are needed to determine all curvature components. The reference body is denoted by the black circle. Note that with the standard deviation equation all ${}^{(1 \dots 6)}u^{a}$, but only ${}^{(1 \dots 3)}\eta^{a}$ are needed in the solution.}
\end{figure}

There are still 3 components of the curvature tensor missing. To determine them, we notice that the following relation between the remaining equations is at our disposal:

\begin{eqnarray}
R_{0312}-R_{0231} &=& \frac{3}{4} {}^{(2,2)}A_3 c^{-1}_{20} c^{-1}_{21} - R_{3020} c_{20} c^{-1}_{21} \nonumber \\ 
&& - R_{3121} c_{21} c^{-1}_{20}, \label{difference_1} \\
R_{0231}-R_{0123} &=& \frac{3}{4} {}^{(4,1)}A_2 c^{-1}_{40} c^{-1}_{43} - R_{2010} c_{40} c^{-1}_{43} \nonumber \\ 
&& - R_{2313} c_{43} c^{-1}_{40}. \label{difference_2}
\end{eqnarray}
Subtracting (\ref{difference_1}) from (\ref{difference_2}) and using the Ricci identity we find:
\begin{eqnarray}
18 : R_{0231} &=& \frac{1}{4} {}^{(4,1)}A_2 c^{-1}_{40} c^{-1}_{43}  -  \frac{1}{4} {}^{(2,2)}A_3 c^{-1}_{20} c^{-1}_{21} \nonumber \\ 
&&  + \frac{1}{3} \big( R_{3020} c_{20} c^{-1}_{21} + R_{3121} c_{21} c^{-1}_{20} \nonumber \\
&&  -  R_{2010} c_{40} c^{-1}_{43} -  R_{2313} c_{43} c^{-1}_{40} \big),  \label{ex0231} \\ 
19 : R_{0312} &=& \frac{1}{4} {}^{(4,1)}A_2 c^{-1}_{40} c^{-1}_{42}  + \frac{1}{2} {}^{(2,2)}A_3 c^{-1}_{20} c^{-1}_{21} \nonumber \\ 
&&  - \frac{1}{3} \big( 2 R_{3020} c_{20} c^{-1}_{21} + 2 R_{3121} c_{21} c^{-1}_{20} \nonumber \\
&&  + R_{2010} c_{40} c^{-1}_{43} +  R_{2313} c_{43} c^{-1}_{40}  \big).  \label{ex0312}
\end{eqnarray}
Finally, by reinsertion of (\ref{difference_1}) in one of the remaining compass equations, one obtains:
\begin{eqnarray}
20 : R_{3212} &=& \frac{3}{4} {}^{(4,1)}A_3  c^{-1}_{20}  c^{-1}_{21}  c_{50}  c^{-1}_{52}  -  \frac{3}{4} {}^{(5,2)}A_3  c^{-1}_{51}  c^{-1}_{52}  \nonumber \\ 
&& + R_{3121}  c^{-1}_{52} \left(c_{51} - c_{50}  c_{21} c^{-1}_{20}\right)  + R_{3220} c_{50} c^{-1}_{51} \nonumber \\
&&  + R_{3020}  c_{50}  c^{-1}_{52} \left(c_{50} c^{-1}_{51} -c_{20}  c^{-1}_{21} \right) .  \label{ex3212}
\end{eqnarray}
By examination of the components given in (\ref{exR1010})-(\ref{ex3212}), we conclude that for a full determination of the curvature one needs 13 test bodies, see fig.\ \ref{fig_standard_compass} for a sketch of the solution. 

\subsubsection{Vacuum spacetime}

\begin{figure}
\begin{center}
\includegraphics[width=7cm,angle=-90]{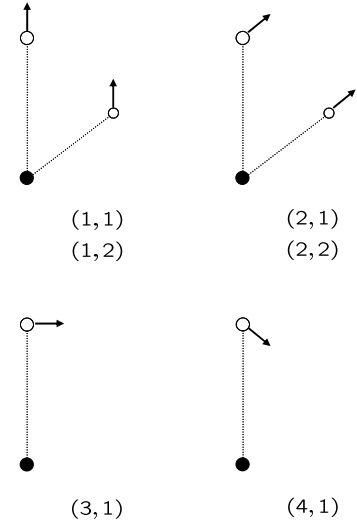}
\end{center}
\caption{\label{fig_standard_compass_vacuum} Symbolical sketch of the explicit compass solution in (\ref{exC1010})-(\ref{exC0312}) for the vacuum case. In total 6 suitably prepared test bodies (hollow circles) are needed to determine all components of the Weyl tensor. The reference body is denoted by the black circle. Note that in vacuum, with the standard deviation equation, all ${}^{(1 \dots 4)}u^{a}$, but only ${}^{(1 \dots 2)}\eta^{a}$ are needed in the solution.}
\end{figure}

In vacuum the number of independent components of the curvature is reduced to the 10 components of the Weyl tensor $C_{abcd}$. Replacing $R_{abcd}$ in the compass solution (\ref{exR1010})-(\ref{ex3212}), and taking into account the symmetries of Weyl we may use a reduced compass setup to completely determine the gravitational field, i.e.\  

\begin{eqnarray}
01 : C_{1010} &=& \frac{3}{4} {}^{(1,1)}A_1 c^{-2}_{10}, \label{exC1010} \\
02 : C_{2010} &=& \frac{3}{4} {}^{(1,1)}A_2 c^{-2}_{10}, \label{exC2010} \\
03 : C_{3010} &=& \frac{3}{4} {}^{(1,1)}A_3 c^{-2}_{10}, \label{exC3010} \\
04 : C_{2020} &=& \frac{3}{4} {}^{(1,2)}A_2 c^{-2}_{10}, \label{exC2020} \\
05 : C_{3020} &=& \frac{3}{4} {}^{(1,2)}A_3 c^{-2}_{10}, \label{exC3020} \\
06 : C_{2110} &=& \frac{3}{4} {}^{(2,1)}A_2 c^{-1}_{21} c^{-1}_{20} -  C_{2010} c^{-1}_{21} c_{20} , \label{exC2110} \\
07 : C_{3110} &=& \frac{3}{4} {}^{(2,1)}A_3 c^{-1}_{21} c^{-1}_{20} -  C_{3010} c^{-1}_{21} c_{20} , \label{exC3110} \\
08 : C_{0212} &=& \frac{3}{4} {}^{(3,1)}A_0 c^{-2}_{32} +  C_{2010} c^{-1}_{32} c_{30} , \label{exC0212} \\
09 : C_{0231} &=& \frac{1}{4} {}^{(4,1)}A_2 c^{-1}_{40} c^{-1}_{43}  -  \frac{1}{4} {}^{(2,2)}A_3 c^{-1}_{20} c^{-1}_{21} \nonumber \\ 
&&  + \frac{1}{3} C_{3020} \big(  c_{20} c^{-1}_{21} + c_{21} c^{-1}_{20}\big) \nonumber \\
&&  - \frac{1}{3} C_{2010}\big( c_{40} c^{-1}_{43} + c_{43} c^{-1}_{40} \big),  \label{exC0231} \\ 
10 : C_{0312} &=& \frac{1}{4} {}^{(4,1)}A_2 c^{-1}_{40} c^{-1}_{42}  + \frac{1}{2} {}^{(2,2)}A_3 c^{-1}_{20} c^{-1}_{21} \nonumber \\ 
&&  - \frac{2}{3} C_{3020} \big( c_{20} c^{-1}_{21} + c_{21} c^{-1}_{20} \big) \nonumber \\
&&  + \frac{1}{3} C_{2010} \big( c_{40} c^{-1}_{43} + c_{43} c^{-1}_{40} \big). \label{exC0312}
\end{eqnarray}
All the other components of the Weyl tensor are obtained from the above by making use of the double-self-duality property $C_{abcd} = -\frac{1}{4}\epsilon_{abef}\epsilon_{cdgh}C^{efgh}$, where $\epsilon_{abcd}$ is the totally antisymmetric Levi-Civita tensor with $\epsilon_{0123}=1$, and the Ricci identity. See fig.\ \ref{fig_standard_compass_vacuum} for a sketch of the  solution.

\subsection{Generalized deviation equation}\label{sec_compass_with_gen_dev}

Let us come back to the generalized deviation equation derived in the first part of the work. In particular the generalized equation with synchronous parametrization for geodesic curves, i.e.\ (\ref{gendev_syn_geo}). Considering this equation at first order, one interesting question is whether it allows for a determination of the curvature with a smaller number of test bodies than the standard deviation equation considered in section \ref{sec_compass_with_standard_dev}.

Rewriting (\ref{gendev_syn_geo}) at first order in normal coordinates yields:
\begin{eqnarray}
\frac{d^2}{ds^2}\eta_a &\stackrel{|_Y}{=}& \frac{4}{3} R_{abcd} u^b \eta^c u^d + 2 R_{abcd} \frac{d\eta^b}{ds}  \eta^c u^d. \label{gen_dev_compass_lower}
\end{eqnarray} 
The apparent difference w.r.t.\ to (\ref{compass_lower}) is that now the velocities of the individual test bodies come into play. 

\subsection{Generalized compass setup}\label{sec_compass_setup_gen_dev}

The lhs of (\ref{gen_dev_compass_lower}) are the measured accelerations and in the following we refer to them by ${}^{(m,n,p)}A_a$. In other words, for $(m)$ compasses, with ${}^{(m)}u^a$ velocities, and $(n)$ test bodies, which can move individually with $(p)$ velocities in one compass, we have the following set of equations:
\begin{eqnarray}
{}^{(m,n,p)}A_a &\stackrel{|_Y}{=}& \frac{4}{3} R_{abcd} \, {}^{(m)}u^b \, {}^{(n)}\eta^c \, {}^{(m)}u^d \nonumber \\
&&+ 2 R_{abcd} \, {}^{(p)}\!\stackrel{\circ}{\eta}\!{}^b  \,  {}^{(n)}\eta^c  \, {}^{(m)}u^d. \label{gen_dev_compass_lower_setup}
\end{eqnarray}  
Here we used the shortcut notation ``$\stackrel{\circ}{\eta}\!{}^a$''$=d\eta^a/ds $ for the standard total derivative. What remains to be chosen, apart from the $(n=1 \dots 5)$ positions of bodies in one compass, the $(m=1 \dots 6)$ actual directions in which each compass / observer shall move, are the $(p=0 \dots 6)$ individual velocities of the bodies. Of course in the end we want to minimize all three numbers, i.e.\ $(m)$, $(n)$, and $(p)$ which are needed to determine all curvature components.  
\begin{eqnarray}
&&{}^{(1)}\!\stackrel{\circ}{\eta}\!{}^a=\left(\begin{array}{c} d_{10} \\ 0 \\ 0\\ 0\\ \end{array} \right),
{}^{(2)}\!\stackrel{\circ}{\eta}\!{}^a=\left(\begin{array}{c}  d_{20} \\ d_{21} \\ 0 \\ 0\\ \end{array} \right),
{}^{(3)}\!\stackrel{\circ}{\eta}\!{}^a=\left(\begin{array}{c}  d_{30} \\ 0 \\ d_{32} \\ 0 \\ \end{array} \right), \nonumber \\
&&{}^{(4)}\!\stackrel{\circ}{\eta}\!{}^a=\left(\begin{array}{c}  d_{40} \\ 0 \\ 0 \\ d_{43}\\ \end{array} \right),
{}^{(5)}\!\stackrel{\circ}{\eta}\!{}^a=\left(\begin{array}{c}  d_{50} \\ d_{51} \\ d_{52} \\ 0\\ \end{array} \right),
{}^{(6)}\!\stackrel{\circ}{\eta}\!{}^a=\left(\begin{array}{c}  d_{60} \\ 0 \\ d_{62} \\ d_{63} \\ \end{array} \right). \nonumber \\
 \label{particle_velocity_setup}
\end{eqnarray}
The $d_{(p)a}$ here are just constants, chosen appropriately to ensure the normalization of the 4-velocity of each test body. Note that in order to recover the results from the previous compass setup in the context of the standard deviation equation, we have just have to choose 
\begin{eqnarray}
&&{}^{(0)}\!\stackrel{\circ}{\eta}\!{}^a=\left(\begin{array}{c}  0 \\ 0 \\ 0 \\ 0 \\ \end{array} \right). \label{zeroth_vel}
\end{eqnarray}

\subsubsection{Explicit curvature components}

\begin{figure}
\begin{center}
\includegraphics[width=7cm,angle=-90]{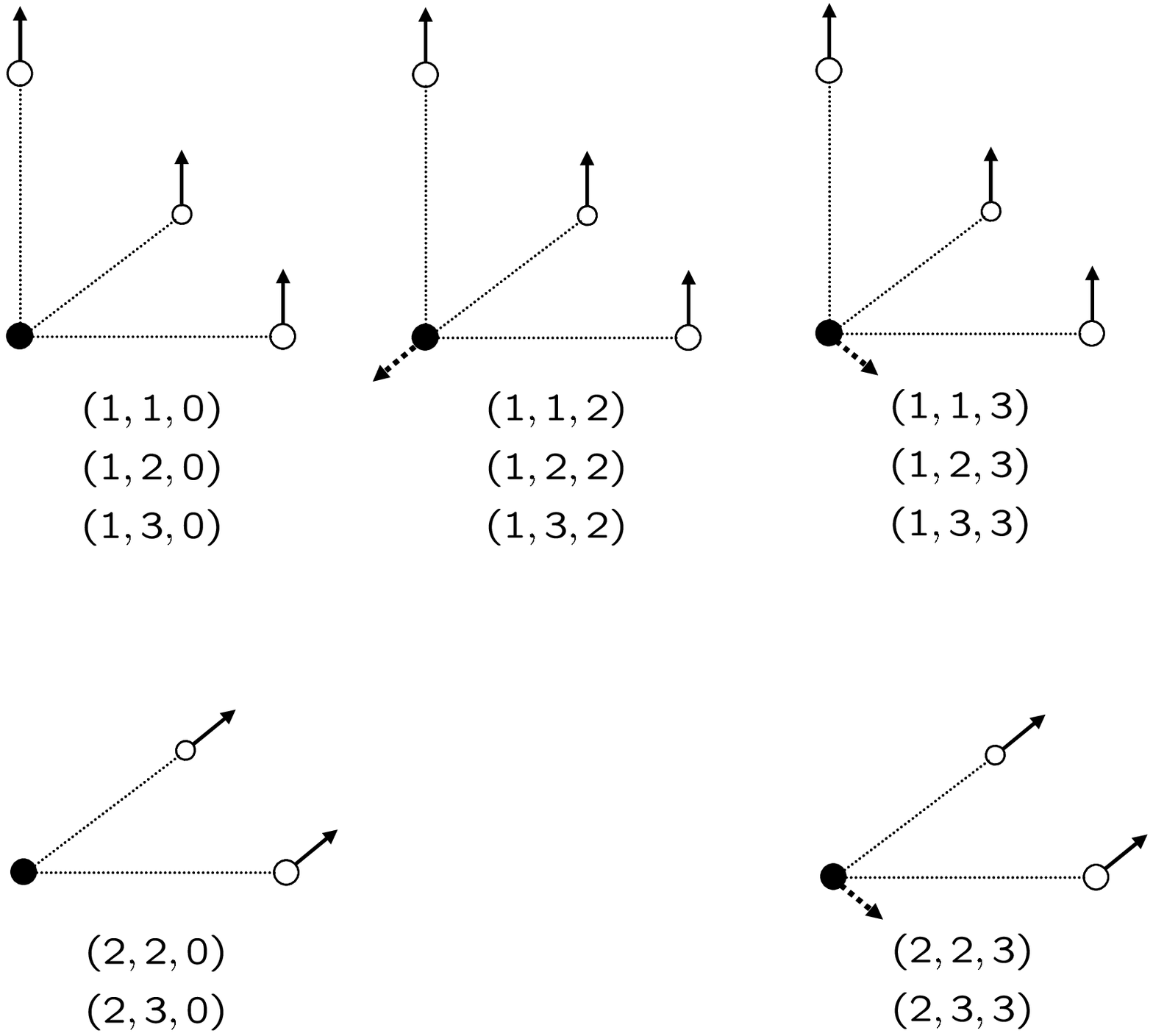}
\end{center}
\caption{\label{fig_generalized_compass} Symbolical sketch of the explicit compass solution in (\ref{gexR1010})-(\ref{gexR3231}). Again, in total 13 suitably prepared test bodies (hollow circles) are needed to determine all curvature components. The reference body is denoted by the black circle. Note that with the generalized deviation equation only ${}^{(1 \dots 2)}u^{a}$, ${}^{(1 \dots 3)}\eta^{a}$, and ${}^{(2 \dots 3)}\!\stackrel{\circ}{\eta}\!{}^a$ are needed in the solution.}
\end{figure}

Similarly to Sec.~\ref{sec_compass_setup_standard_dev}, the 20 independent components of the curvature tensor can be explicitly determined in terms of the accelerations ${}^{(m,n,p)}A_a$ and velocities ${}^{(m)}u^a$ via the deviation equation (\ref{gen_dev_compass_lower_setup}) by using the compass configuration given in (\ref{position_setup}), (\ref{velocity_setup}), and (\ref{particle_velocity_setup}). The result reads as follows
\begin{widetext}
\begin{eqnarray}
01 : R_{1010} &=& \frac{3}{4} {}^{(1,1,0)}A_1 c^{-2}_{10}, \label{gexR1010} \\
02 : R_{2010} &=& \frac{3}{4} {}^{(1,1,0)}A_2 c^{-2}_{10}, \label{gexR2010} \\
03 : R_{3010} &=& \frac{3}{4} {}^{(1,1,0)}A_3 c^{-2}_{10}, \label{gexR3010} \\
04 : R_{2110} &=& \frac{1}{2} {}^{(1,1,2)}A_2 d^{-1}_{21} c^{-1}_{10} - \left(\frac{2}{3} d^{-1}_{21}  c_{10} + d_{20} d^{-1}_{21} \right) R_{2010},\label{gexR2110} \\
05 : R_{3110} &=& \frac{1}{2} {}^{(1,1,2)}A_3 d^{-1}_{21} c^{-1}_{10} - \left(\frac{2}{3} d^{-1}_{21}  c_{10} + d_{20} d^{-1}_{21} \right) R_{3010},\label{gexR3110} \\
06 : R_{3210} &=& \frac{1}{2} {}^{(1,1,3)}A_3 d^{-1}_{32} c^{-1}_{10} - \left(\frac{2}{3} d^{-1}_{32}  c_{10} + d_{30} d^{-1}_{32} \right) R_{3010},\label{gexR3210} \\
07 : R_{2020} &=& \frac{3}{4} {}^{(1,2,0)}A_2 c^{-2}_{10}, \label{gexR2020} \\
08 : R_{3020} &=& \frac{3}{4} {}^{(1,2,0)}A_3 c^{-2}_{10}, \label{gexR3020} \\
09 : R_{2120} &=& \frac{1}{2} {}^{(1,2,2)}A_2 d^{-1}_{21} c^{-1}_{10} - \left(\frac{2}{3} d^{-1}_{21}  c_{10} + d_{20} d^{-1}_{21} \right) R_{2020},\label{gexR2120} \\
10 : R_{3120} &=& \frac{1}{2} {}^{(1,2,2)}A_3 d^{-1}_{21} c^{-1}_{10} - \left(\frac{2}{3} d^{-1}_{21}  c_{10} + d_{20} d^{-1}_{21} \right) R_{3020},\label{gexR3120} \\
11 : R_{3220} &=& \frac{1}{2} {}^{(1,2,3)}A_3 d^{-1}_{32} c^{-1}_{10} - \left(\frac{2}{3} d^{-1}_{32}  c_{10} + d_{30} d^{-1}_{32} \right) R_{3020},\label{gexR3220} 
\end{eqnarray}
\end{widetext}
\begin{widetext}
\begin{eqnarray}
12 : R_{3030} &=& \frac{3}{4} {}^{(1,3,0)}A_3 c^{-2}_{10}, \label{gexR3030}\\
13 : R_{2130} &=& \frac{1}{2} {}^{(1,3,2)}A_2 d^{-1}_{21} c^{-1}_{10} - \left(\frac{2}{3} d^{-1}_{21}  c_{10} + d_{20} d^{-1}_{21} \right) R_{2030},\label{gexR2130}\\
14 : R_{3130} &=& \frac{1}{2} {}^{(1,3,2)}A_3 d^{-1}_{21} c^{-1}_{10} - \left(\frac{2}{3} d^{-1}_{21}  c_{10} + d_{20} d^{-1}_{21} \right) R_{3030},\label{gexR3130} \\
15 : R_{3230} &=& \frac{1}{2} {}^{(1,3,3)}A_3 d^{-1}_{32} c^{-1}_{10} - \left(\frac{2}{3} d^{-1}_{32}  c_{10} + d_{30} d^{-1}_{32} \right) R_{3030},\label{gexR3230} \\
16 : R_{2121} &=& \frac{3}{4} {}^{(2,2,0)}A_2 c^{-2}_{21} -  R_{2020} c^{2}_{20} c^{-2}_{21} - 2 R_{2120} c_{20} c^{-1}_{21},\label{gexR2121} \\
17 : R_{3121} &=& \frac{3}{4} {}^{(2,2,0)}A_3 c^{-2}_{21} -  R_{3020} c^{2}_{20} c^{-2}_{21} - \left( R_{3021} + R_{3120}\right) c_{20} c^{-1}_{21},\label{gexR3121} \\
18 : R_{3221} &=& \frac{1}{2} {}^{(2,2,3)}A_3 d^{-1}_{32} c^{-1}_{21} -  \left(\frac{2}{3} d^{-1}_{32} c_{20}+ d_{30} d^{-1}_{32} \right) R_{3021} -  \left(\frac{2}{3} d^{-1}_{32} c^{-1}_{21} c^2_{20} + d^{-1}_{32} d_{30} c^{-1}_{21} c_{20} \right) R_{3020} \nonumber \\ 
&& - \frac{2}{3} R_{3120} d^{-1}_{32} c_{20} - \frac{2}{3} R_{3121} d^{-1}_{32} c_{21} - R_{3220} c^{-1}_{21} c_{20} ,\label{gexR3221} \\
19 : R_{3131} &=& \frac{3}{4} {}^{(2,3,0)}A_3 c^{-2}_{21} -  R_{3030} c_{20} c^{-2}_{21} - 2 R_{3130} c_{20} c^{-1}_{21},\label{gexR3131} \\
20 : R_{3231} &=& \frac{1}{2} {}^{(2,3,3)}A_3 d^{-1}_{32} c^{-1}_{21} - \left( \frac{4}{3} d^{-1}_{32} c_{20} + d^{-1}_{32} d_{30} \right)  R_{3031} - \left(\frac{2}{3} d^{-1}_{32} c^{-1}_{21} c^2_{20} + d^{-1}_{32} d_{30} c^{-1}_{21} c_{20} \right) R_{3030} \nonumber\\ && - \frac{2}{3} d^{-1}_{32} c_{21}  R_{3131}  - R_{3230} c^{-1}_{21} c_{20} ,\label{gexR3231}
\end{eqnarray}
\end{widetext}
By examination of the components given in (\ref{gexR1010})-(\ref{gexR3231}), we infer that for a full determination of the curvature by means of the generalized deviation equation one again needs 13 test bodies. See fig.\ \ref{fig_generalized_compass} for a sketch of the solution. 

\subsubsection{Vacuum spacetime}

\begin{table}
\caption{\label{tab_no_bodies}Number of required bodies in a compass setup for a full determination of the gravitational field, i.e.\ 20 components of $R_{abcd}$ in a general spacetime, and 10 components of $C_{abcd}$ in vacuum.}
\begin{ruledtabular}
\begin{tabular}{lll}
&\multicolumn{2}{c}{{Spacetime}}\\
 & General & Vacuum \\
\hline
&&\\
Standard deviation equation&13&6\\
&&\\
Generalized deviation equation&13&5\\
&&\\
\end{tabular}
\end{ruledtabular}
\end{table}

By replacing $R_{abcd}$ in the compass solution (\ref{gexR1010})-(\ref{gexR3231}), and taking into account the symmetries of Weyl we may use a reduced compass setup to completely determine the gravitational field, this time by means of the generalized deviation equation. Explicitly, one ends up with
\begin{eqnarray}
01 : C_{1010} &=& \frac{3}{4} {}^{(1,1,0)}A_1 c^{-2}_{10}, \label{gexC1010} \\
02 : C_{2010} &=& \frac{3}{4} {}^{(1,1,0)}A_2 c^{-2}_{10}, \label{gexC2010} \\
03 : C_{3010} &=& \frac{3}{4} {}^{(1,1,0)}A_3 c^{-2}_{10}, \label{gexC3010} \\
04 : C_{2110} &=& \frac{1}{2} {}^{(1,1,2)}A_2 d^{-1}_{21} c^{-1}_{10} \nonumber \\ && 
- \left(\frac{2}{3} d^{-1}_{21}  c_{10} + d_{20} d^{-1}_{21} \right) C_{2010},\label{gexC2110} \\
05 : C_{3110} &=& \frac{1}{2} {}^{(1,1,2)}A_3 d^{-1}_{21} c^{-1}_{10} \nonumber \\ &&
- \left(\frac{2}{3} d^{-1}_{21}  c_{10} + d_{20} d^{-1}_{21} \right) C_{3010},\label{gexC3110} \\
06 : C_{3210} &=& \frac{1}{2} {}^{(1,1,3)}A_3 d^{-1}_{32} c^{-1}_{10} \nonumber \\ &&
- \left(\frac{2}{3} d^{-1}_{32}  c_{10} + d_{30} d^{-1}_{32} \right) C_{3010},\label{gexC3210} \\
07 : C_{2020} &=& \frac{3}{4} {}^{(1,2,0)}A_2 c^{-2}_{10}, \label{gexC2020} \\
08 : C_{3020} &=& \frac{3}{4} {}^{(1,2,0)}A_3 c^{-2}_{10}, \label{gexC3020} \\
09 : C_{2120} &=& \frac{1}{2} {}^{(1,2,2)}A_2 d^{-1}_{21} c^{-1}_{10} \nonumber \\ && 
- \left(\frac{2}{3} d^{-1}_{21}  c_{10} + d_{20} d^{-1}_{21} \right) C_{2020},\label{gexC2120} \\
10 : C_{3120} &=& \frac{1}{2} {}^{(1,2,2)}A_3 d^{-1}_{21} c^{-1}_{10} \nonumber \\ && 
- \left(\frac{2}{3} d^{-1}_{21}  c_{10} + d_{20} d^{-1}_{21} \right) C_{3020}.\label{gexC3120}
\end{eqnarray}

\begin{figure}
\begin{center}
\includegraphics[width=3.5cm,angle=-90]{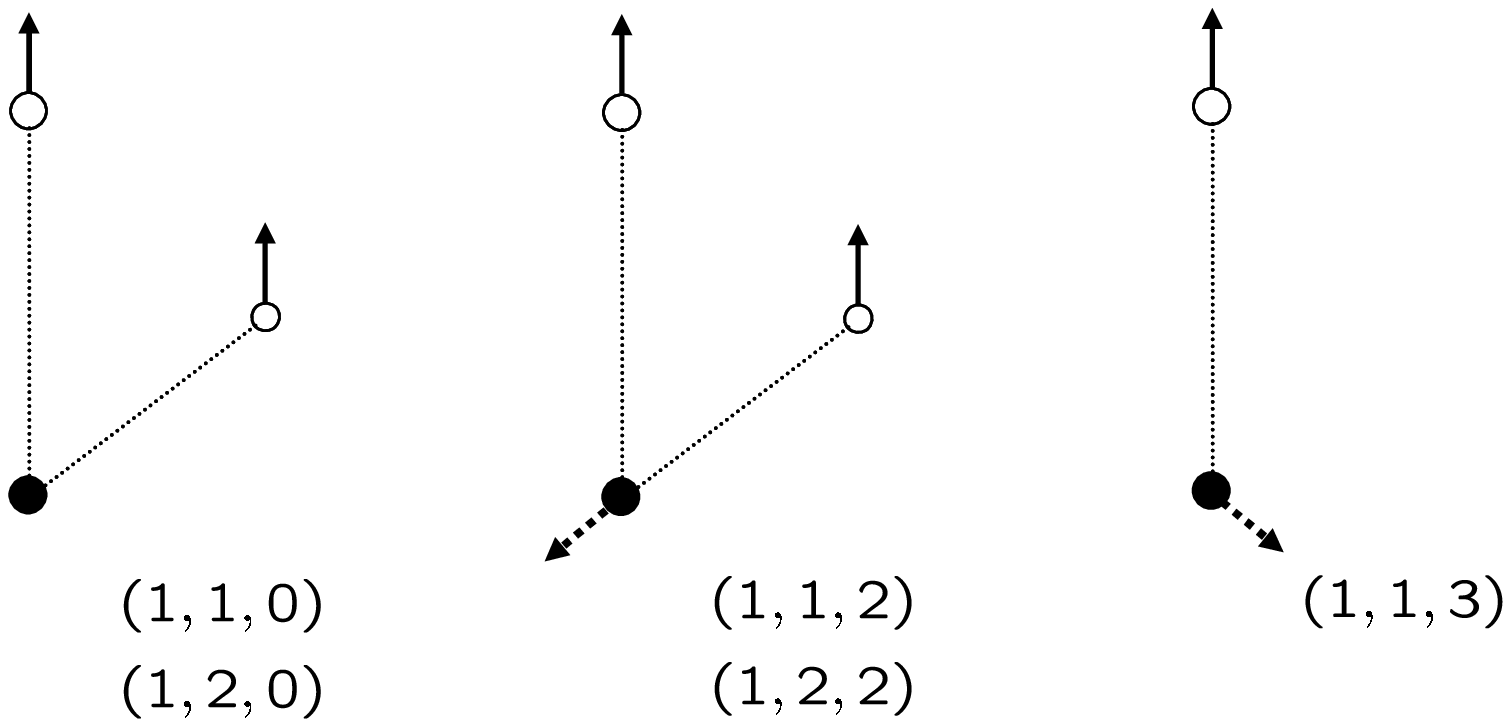}
\end{center}
\caption{\label{fig_generalized_compass_vacuum} Symbolical sketch of the explicit compass solution in (\ref{gexC1010})-(\ref{gexC3120}) for the vacuum case. In total 5 suitably prepared test bodies (hollow circles) are needed to determine all components of the Weyl tensor. The reference body is denoted by the black circle. Note that in vacuum, with the generalized deviation equation, only ${}^{(1)}u^{a}$, ${}^{(1 \dots 2)}\eta^{a}$, and ${}^{(2 \dots 3)}\!\stackrel{\circ}{\eta}\!{}^a$ are needed in the solution.}
\end{figure}

See fig.\ \ref{fig_generalized_compass_vacuum} for a sketch of the solution. Table \ref{tab_no_bodies} contains an overview of the number of required test bodies in different compass setups and in general as well as in vacuum spacetimes.

\section{Conclusion}\label{sec_conclusions}

In this work we derived a generalized covariant deviation equation in the framework of Synge's world function approach. It should be stressed that our exact deviation equation in (\ref{eta_2nd_deriv_alternative_2}) is valid for arbitrary world lines and in general background spacetimes. In the subsequent analysis we provided a systematic expansion of the exact deviation equation up to the third order in the world function (\ref{eta_2nd_deriv_compact_2}). This equation can be viewed as a generalization of the well-known geodesic deviation equation, to which it was shown to reduce under the right assumptions. As we have shown in detail, our results encompass several suggestions for a generalized deviation equation from the literature as special cases, and therefore may serve a unified framework for further studies.  

In a subsequent analysis we have shown how deviation equations can be used to determine the curvature of spacetime. For this we extended the notion of a gravitational compass \cite{Szekeres:1965} and worked out compass setups for general as well as for vacuum spacetimes. One setup is based on the standard geodesic deviation equation (\ref{geodesic_dev_revovered_affine}), and another is based on the next order generalization given in (\ref{gendev_syn_geo}) which goes beyond the linearized case. For both cases we provided the explicit compass solution which allows for a full determination of the curvature. 

In contrast to the general considerations in \cite{Synge:1960,Szekeres:1965} we give an explicit exact solution for the compass setup. With the standard deviation equation, as well as with the generalized deviation equation, we need at least 13 test bodies to determine all curvature components in a general spacetime. For the standard deviation we therefore obtain the same number of bodies as in \cite{Ciufolini:Demianski:1986}, however it is worthwhile to note that no explicit solution was given in \cite{Ciufolini:Demianski:1986} for a non-vacuum spacetime. In the case of a generalized deviation equation our findings are at odds with the results in \cite{Ciufolini:Demianski:1986}. However, this discrepancy in the generalized case comes as no surprise since the generalized equation used in \cite{Ciufolini:Demianski:1986} -- which was previously derived in \cite{Ciufolini:1986} -- differs from our equation. In vacuum spacetimes, we have explicitly shown that the number of required test bodies is reduced to 6, for the standard deviation equation, and to 5, for the generalized deviation equation. 

Furthermore, it is interesting to note that in the case of the standard deviation equation, the opinion of the authors \cite{Synge:1960,Szekeres:1965} differs when it comes to the number of required test bodies. This seems to be related to the counting scheme and the interpretation of the notion of a compass. Since no explicit compass solutions were given in \cite{Synge:1960,Szekeres:1965}, one cannot make a comparison to our results. In the case of \cite{Ciufolini:Demianski:1986}, we were not able to verify that the given solution does fulfill the compass equations derived in that work. However, the agreement on the number of required bodies in combination with the standard deviation is reassuring. 

In summary, we have explicitly shown how deviation equations can be used to measure the gravitational field. Our results are of direct operational relevance and form the basis for many experiments. Important applications range from the description of gravitational wave detectors to the study of satellite configurations for gravitational field mapping in relativistic geodesy. An interesting question is whether a further reduction of the number of required test bodies for certain experiments is possible. A systematic analysis of the practical applications of generalized deviation equations, including the gravitation wave detection, will be presented elsewhere.

\begin{acknowledgments}
This work was supported by the Deutsche Forschungsgemeinschaft (DFG) through the grant SFB 1128/1 (D.P.). The work of Y.N.O. was partially supported by PIER (``Partnership for Innovation, Education and Research'' between DESY and Universit\"at Hamburg) and by the Russian Foundation for Basic Research (Grant No. 16-02-00844-A). 
\end{acknowledgments}

\appendix

\section{Notations and conventions}\label{sec_notation}

\begin{table}
\caption{\label{tab_symbols}Directory of symbols.}
\begin{ruledtabular}
\begin{tabular}{ll}
Symbol & Explanation\\
\hline
&\\
\hline
\multicolumn{2}{l}{{Geometrical quantities}}\\
\hline
$g_{a b}$ & Metric\\
$\sqrt{-g}$ & Determinant of the metric \\
$\delta^a_b$ & Kronecker symbol \\
$\epsilon_{abcd} $ & Levi-Civita symbol\\
$x^{a}$, $s$ & Coordinates, proper time \\
$\Gamma_{a b}{}^c$ & Connection \\
${}^{\ast}\Gamma_{a b \dots}{}^c$ & Deriv. conn. (normal coords.)\\
$R_{a b c}{}^d$, $C_{a b c}{}^d$ & Riemann, Weyl curvature \\
$\sigma$ & World function\\
$\eta^y$ & Deviation vector\\
$g^{y_0}{}_{x_0}$ & Parallel propagator\\
&\\
\hline
\multicolumn{2}{l}{{Misc}}\\
\hline
$Y(t), X(t)$ & (Reference) world line\\
$u^a$ & Velocity \\
$a^b$ & Acceleration \\
$K^{x}{}_{y}, H^{x}{}_{y}$ & Jacobi propagators \\
${}^{(m,n)}A_a$, ${}^{(m,n,p)}A_a$ & Accelerations of \\
& compass constituents \\
&\\
\hline
\multicolumn{2}{l}{{Auxiliary quantities}}\\
\hline
$\alpha^{y_0}{}_{y_1 \dots y_n}$, $\beta^{y_0}{}_{y_1 \dots y_n}$, $\gamma^{y_0}{}_{y_1 \dots y_n}$& Expansion coefficients\\
$c_{(m)a}$, $d_{(m)a}$& Constants\\
$\phi^{y_1}{}_{y_2 \dots}$, $\lambda^{y_1}{}_{y_2 \dots}$, $\mu^{y_1}{}_{y_2 \dots}$, & Abbreviations \\
$\Delta_{i_1\dots}$, $\Xi_{i_1\dots}$ &\\
&\\
\hline
\multicolumn{2}{l}{{Operators}}\\
\hline
$\partial_i$, $\nabla_i$ & (Partial, covariant) derivative \\ 
$\frac{D}{ds} = $``$\dot{\phantom{a}}$'' & Total cov. derivative \\
$\frac{d}{ds} = $``$\stackrel{\circ}{\phantom{a}}$'' & Total  derivative \\
``$[ \dots ]$''& Coincidence limit\\
&\\
\end{tabular}
\end{ruledtabular}
\end{table}

Our conventions for the Riemann curvature are as follows:
\begin{eqnarray}
&& 2 T^{c_1 \dots c_k}{}_{d_1 \dots d_l ; [ba] } \equiv 2 \nabla_{[a} \nabla_{b]} T^{c_1 \dots c_k}{}_{d_1 \dots d_l} \nonumber \\
& = & \sum^{k}_{i=1} R_{abe}{}^{c_i} T^{c_1 \dots e \dots c_k}{}_{d_1 \dots d_l} \nonumber \\
&& - \sum^{l}_{j=1} R_{abd_j}{}^{e} T^{c_1 \dots c_k}{}_{d_1 \dots e \dots d_l}. \label{curvature_def}
\end{eqnarray}
The Ricci tensor is introduced by $R_{ij} = R_{kij}{}^k$, and the curvature scalar is $R = g^{ij}R_{ij}$. The signature of the spacetime metric is assumed to be $(+1,-1,-1,-1)$.

In the following, we summarize some of the frequently used formulas in the context of the bitensor formalism [in particular for the world function $\sigma(x,y)$], see, e.g., \cite{Synge:1960,DeWitt:Brehme:1960,Poisson:etal:2011} for the corresponding derivations. Note that our curvature conventions differ from those in \cite{Synge:1960,Poisson:etal:2011}. Indices attached to the world function always denote covariant derivatives, at the given point, i.e.\ $\sigma_y:= \nabla_y \sigma$; hence, we do not make explicit use of the semicolon in case of the world function. We start by stating, without proof, the following useful rule for a bitensor $B$ with arbitrary indices at different points (here just denoted by dots):
\begin{eqnarray}
\left[B_{\dots} \right]_{;y} = \left[B_{\dots ; y} \right] + \left[B_{\dots ; x} \right]. \label{synges_rule}
\end{eqnarray}
Here a coincidence limit of a bitensor $B_{\dots}(x,y)$ is a tensor 
\begin{eqnarray}
\left[B_{\dots} \right] = \lim\limits_{x\rightarrow y}\,B_{\dots}(x,y),\label{coin}
\end{eqnarray}
determined at $y$. Furthermore, we collect the following useful identities: 
\begin{eqnarray}
&&\sigma_{y_0 y_1 x_0 y_2 x_1} = \sigma_{y_0 y_1 y_2 x_0 x_1} = \sigma_{x_0 x_1 y_0 y_1 y_2 }, \label{rule_1} \\
&&g^{x_1 x_2} \sigma_{x_1} \sigma_{x_2} = 2 \sigma = g^{y_1 y_2} \sigma_{y_1} \sigma_{y_2}, \label{rule_2}\\
&&\left[ \sigma \right]=0, \quad  \left[ \sigma_x \right] = \left[ \sigma_y \right]  = 0, \label{rule_3} \\
&& \left[ \sigma_{x_1 x_2} \right] =  \left[ \sigma_{y_1 y_2} \right] = g_{y_1 y_2}, \label{rule_4}\\ 
&& \left[ \sigma_{x_1 y_2} \right] =  \left[ \sigma_{y_1 x_2} \right] = - g_{y_1 y_2}, \label{rule_5}\\ 
&& \left[ \sigma_{x_1 x_2 x_3} \right] = \left[ \sigma_{x_1 x_2 y_3} \right] = \left[ \sigma_{x_1 y_2 y_3} \right] = \left[ \sigma_{y_1 y_2 y_3} \right] = 0, \nonumber \\ \label{rule_6}\\
&&\left[g^{x_0}{}_{y_1} \right] = \delta^{y_0}{}_{y_1}, \quad \left[g^{x_0}{}_{y_1 ; x_2} \right] = \left[g^{x_0}{}_{y_1 ; y_2} \right] = 0, \label{rule_7} \\
&& \left[g^{x_0}{}_{y_1 ; x_2 x_3} \right] = \frac{1}{2} R^{y_0}{}_{y_1 y_2 y_3}. \label{rule_8}
\end{eqnarray}

\section{Normal coordinates}\label{sec_normal_coordinates}

Here we provide the explicit expressions of the derivatives of the Riemannian connection $\Gamma_{i_1\dots i_Nij}{}^k := \partial_{i_1}\dots\partial_{i_N}\Gamma_{ij}{}^k$ in normal coordinates.

The list of the lowest derivatives for $N = 0,1,2,3$ reads as follows:
\begin{eqnarray}
{}^{\ast}\Gamma_{ij}{}^k &=& 0,\label{gamma0}\\
{}^{\ast}\Gamma_{i_1ij}{}^k &=& {\frac 23}\,R_{i_1(ij)}{}^k,\label{gamma1}\\
{}^{\ast}\Gamma_{i_1i_2ij}{}^k &=& {\frac 16}\left[5\nabla_{(i_1}R_{i_2)(ij)}{}^k - \nabla_{(i}R_{j)(i_1i_2)}{}^k\right],\label{gamma2}\\
{}^{\ast}\Gamma_{i_1i_2i_3ij}{}^k &=& {\frac 3{20}}\left[6\nabla_{(i_1}\nabla_{i_2} R_{i_3)(ij)}{}^k - \nabla_i\nabla_{(i_1}R_{|j|i_2i_3)}{}^k \right. \nonumber \\ 
&& \left. - \nabla_j\nabla_{(i_1}R_{|i|i_2i_3)}{}^k\right]+{\frac{4}{15}}\,R_{p(i_1i_2}{}^k\,R_{i_3)(ij)}{}^p\nonumber\\
&& + \frac{1}{2}\left[R_{p(i_1i_2}{}^k\,R_{i_3)(ij)}{}^p - R_{i(i_1i_2}{}^p\,R_{i_3)(jp)}{}^k \right. \nonumber \\ 
&& \left. - R_{j(i_1i_2}{}^p\,R_{i_3)(ip)}{}^k \right] + \frac{1}{10} \left[R_{i(i_1i_2}{}^p\,R_{i_3)jp}{}^k \right. \nonumber \\ 
&& \left. + R_{j(i_1i_2}{}^p\,R_{i_3)ip}{}^k\right].\label{gamma3}
\end{eqnarray}
The parentheses denote the symmetrization over the enclosed indices; indices between the vertical lines are excluded from the symmetrization. As a check, from these formulas we can derive the symmetrized derivatives of the connection which are better known in the literature (see, e.g., Petrov \cite{Petrov:1969}):
\begin{eqnarray}
{}^{\ast}\Gamma_{(i_1i)j}{}^k &=& -\,{\frac{1}{3}}\,R_{j(i_1i)}{}^k,\label{gammaS1}\\
{}^{\ast}\Gamma_{(i_1i_2i)j}{}^k &=& -\,{\frac{1}{2}}\,\nabla_{(i_1}R_{|j|i_2i)}{}^k, \label{gammaS2}\\
{}^{\ast}\Gamma_{(i_1i_2i_3i)j}{}^k &=& -\,{\frac{3}{5}}\nabla_{(i_1}\nabla_{i_2} R_{|j|i_3i)}{}^k 
\nonumber \\ && - {\frac{2}{15}}\,R_{p(i_1i_2}{}^k\,R_{|j|i_3i)}{}^p.\label{gammaS3}
\end{eqnarray}
It is worthwhile to note that the symmetrization of the two last lines in (\ref{gamma3}) over $(i_1i_2i_3i)$ yields zero.

The above formulas can be derived as follows. The derivatives of the connection ${}^{\ast}\Gamma_{i_1\cdots i_N\,ij}{}^k$ satisfy the algebraic equations
\begin{equation}
{}^{\ast}\Gamma_{i_1i_2\cdots i_N\,ij}{}^k - {}^{\ast}\Gamma_{ii_2\cdots i_N\,i_1j}{}^k = \Delta_{i_1iji_2\cdots i_N}{}^k,\label{gaga}
\end{equation}
where $\Delta_{i_1iji_2\cdots i_N}{}^k$ are the tensors with the symmetry properties
\begin{eqnarray}
\Delta_{i_1iji_2\cdots i_N}{}^k &=& -\,\Delta_{ii_1ji_2\cdots i_N}{}^k,\\
\Delta_{i_1iji_2\cdots i_N}{}^k &=& \Delta_{ii_1j(i_2\cdots i_N)}{}^k,\\
\Delta_{[i_1ij]i_2\cdots i_N}{}^k &=& 0,\\
\Delta_{[i_1i|j|i_2]\cdots i_N}{}^k &=& 0.
\end{eqnarray}
That is, these tensors are skew-symmetric in the first two indices and totally symmetric in the last $N-1$ indices (these properties are thus consistent with the symmetry properties of the left-hand side of the equation (\ref{gaga})), and in addition, the antisymmetrization over the first three indices and over the first pair and the fourth index vanishes. Using these symmetry properties, one can solve the equation (\ref{gaga}) with respect to the derivatives of the connection. In symbolic form, the general solution (for any $N$) reads
\begin{eqnarray}
{}^{\ast}\Gamma_{i_1i_2\cdots i_N\,ij}{}^k &=& \frac {1}{K}\left[(K-1)\Delta_{\rm perm}{}^k + (K-2)\Delta_{\rm perm}{}^k \right. \nonumber \\
&& \left. + \cdots + \Delta_{\rm perm}{}^k\right],
\end{eqnarray}
where $K = (N+2)(N+1)/2$ and the right-hand side contains $(K-1)$ terms in which the $(N + 2)$ lower indices of $\Delta$'s are permuted in accordance with a certain rule. Actually, the determination of this permutation rule is a highly nontrivial problem which is related to the famous theorem of Desargues, as was shown by Veblen \cite{Veblen:1922}.

We will only give the solutions for the case of $N = 1,2,3,4$:
\begin{eqnarray}
{}^{\ast}\Gamma_{i_1ij}{}^k &=& {\frac{1}{3}}\left[2\Delta_{i_1ji}{}^k + \Delta_{jii_1}{}^k\right],\label{gamma1v}\\
{}^{\ast}\Gamma_{i_1i_2ij}{}^k &=& {\frac{1}{6}}\left[5\Delta_{i_1jii_2}{}^k + 4\Delta_{jii_1i_2}{}^k + 3\Delta_{i_2ji_1i}{}^k \right.\nonumber\\
&&\left. +\,2\Delta_{ii_1i_2j}{}^k + \Delta_{jii_2i_1}{}^k\right],\label{gamma2v}\\
{}^{\ast}\Gamma_{i_1i_2i_3ij}{}^k &=& {\frac{1}{10}}\left[9\Delta_{i_1jii_2i_3}{}^k + 8\Delta_{jii_1i_2i_3}{}^k  \right.\nonumber\\
&& + 7\Delta_{i_2ji_1ii_3}{}^k +\,6 \Delta_{ii_1i_2ji_3}{}^k  \nonumber\\
&& + 5\Delta_{jii_2i_1i_3}{}^k + 4\Delta_{i_3ji_2ii_1}{}^k \nonumber\\
&& +\,3 \Delta_{i_1i_2i_3ji}{}^k + 2\Delta_{ii_1i_3i_2j}{}^k \nonumber \\
&& +\, \left. \Delta_{jii_3i_1i_2}{}^k\right],\label{gamma3v}
\end{eqnarray}
\begin{eqnarray}
{}^{\ast}\Gamma_{i_1i_2i_3i_4ij}{}^k &=& {\frac{1}{15}}\left[14\Delta_{i_1jii_2i_3i_4}{}^k + 13\Delta_{jii_1i_2i_3i_4}{}^k  \right.\nonumber\\
&& +\,12\Delta_{i_2ji_1ii_3i_4}{}^k + 11 \Delta_{ii_1i_2ji_3i_4}{}^k \nonumber\\
&& +\,10\Delta_{jii_2i_1i_3i_4}{}^k + 9\Delta_{i_3ji_2ii_1i_4}{}^k \nonumber \\
&& +\,8\Delta_{i_1i_2i_3jii_4}{}^k + 7\Delta_{ii_1i_3i_2ji_4}{}^k\nonumber\\
&& + 6\Delta_{jii_3i_1i_2i_4}{}^k +\,5\Delta_{i_4ji_3ii_1i_2}{}^k \nonumber\\
&& + 4\Delta_{i_2i_3i_4jii_1}{}^k + 3\Delta_{i_1i_2i_4i_3ji}{}^k\nonumber\\
&& \left. + 2\Delta_{ii_1i_4i_2i_3j}{}^k + \Delta_{jii_4i_1i_2i_3}{}^k \right].\label{gamma4v}
\end{eqnarray}
By differentiating covariantly the curvature tensor $R_{ijl}{}^k$, one can straightforwardly identify the $\Delta$'s with the polynomials built from the curvature and its derivatives. Explicitly, we have
\begin{eqnarray}
\Delta_{i_1ij}{}^k &=& R_{i_1ij}{}^k,\label{beta1}\\
\Delta_{i_1iji_2}{}^k &=& \nabla_{i_2}R_{i_1ij}{}^k,\label{beta2}\\
\Delta_{i_1iji_2i_3}{}^k &=& \nabla_{(i_2}\nabla_{i_3)}R_{i_1ij}{}^k + \Xi_{i_1iji_2i_3}{}^k,\label{beta3}
\end{eqnarray}
where the quadratic in curvature contraction reads
\begin{eqnarray}
\hspace{-0.6cm} \Xi_{i_1iji_2i_3}{}^k &&= {\frac 13}\left[R_{j(i_2i_3)}{}^p\,R_{ii_1p}{}^k - R_{p(i_2i_3)}{}^k\,R_{ii_1j}{}^p\right.\nonumber\\
&& +\left. R_{i(i_2i_3)}{}^p\,R_{pi_1j}{}^k - R_{i_1(i_2i_3)}{}^p\,R_{pij}{}^k\right]\nonumber\\
&& +\,{\frac{4}{9}}\left[R_{i_2(ip)}{}^k\,R_{i_3(i_1j)}{}^p + R_{i_3(ip)}{}^k\,R_{i_2(i_1j)}{}^p \right.\nonumber\\
&& -\left. R_{i_2(i_1p)}{}^k\,R_{i_3(ij)}{}^p - R_{i_3(i_1p)}{}^k\,R_{i_2(ij)}{}^p\right].\label{beta4}
\end{eqnarray}
Inserting (\ref{beta1})-(\ref{beta4}) into (\ref{gamma1v})-(\ref{gamma3v}), we finally obtain the expressions (\ref{gamma1})-(\ref{gamma3}).

It is worthwhile to mention that all the formulas derived here (in accordance with the general theory of normal coordinates \cite{Veblen:1922,Veblen:Thomas:1923,Thomas:1934}) are valid not only for the Riemannian Christoffel symbols but for an arbitrary symmetric connection $\Gamma_{ij}{}^k$ too. The explicit higher order results (\ref{gamma3}), (\ref{gamma3v}), (\ref{gamma4v}) and (\ref{beta4}) are new.

A direct prescription how to calculate the derivatives of the connection is described in \cite{Avramidi:1991,Avramidi:1995}, although it seems impossible to give an explicit general formula. However, using the recursive prescriptions of \cite{Avramidi:1991,Avramidi:1995}, one can find the ${}^{\ast}\Gamma$'s for any $N$.

\bibliographystyle{unsrtnat}
\bibliography{deviation_bibliography}

\begin{thebibliography}{72}
\providecommand{\natexlab}[1]{#1}
\providecommand{\url}[1]{\texttt{#1}}
\expandafter\ifx\csname urlstyle\endcsname\relax
  \providecommand{\doi}[1]{doi: #1}\else
  \providecommand{\doi}{doi: \begingroup \urlstyle{rm}\Url}\fi

\bibitem[{Synge}(1960)]{Synge:1960}
J.~L. {Synge}.
\newblock \emph{{Relativity: The general theory}}.
\newblock North-Holland, Amsterdam, 1960.

\bibitem[{DeWitt} and {Brehme}(1960)]{DeWitt:Brehme:1960}
B.~S. {DeWitt} and R.~W. {Brehme}.
\newblock {Radiation damping in a gravitational field}.
\newblock \emph{Ann. Phys. (N.Y.)}, 9:\penalty0 220, 1960.

\bibitem[{Dixon}(1964)]{Dixon:1964}
W.~G. {Dixon}.
\newblock {A covariant multipole formalism for extended test bodies in General
  Relativity}.
\newblock \emph{Nuovo Cimento}, 34:\penalty0 317, 1964.

\bibitem[{Dixon}(1974)]{Dixon:1974}
W.~G. {Dixon}.
\newblock {Dynamics of extended bodies in General Relativity. III. Equations of
  motion}.
\newblock \emph{Phil. Trans. R. Soc. Lond. A}, 277:\penalty0 59, 1974.

\bibitem[{Dixon}(1979)]{Dixon:1979}
W.~G. {Dixon}.
\newblock {Extended bodies in General Relativity: Their description and
  motion}.
\newblock \emph{Proc. Int. School of Phys. Enrico Fermi LXVII, Ed. J. Ehlers,
  North Holland, Amsterdam}, page 156, 1979.

\bibitem[{Dixon}(2008)]{Dixon:2008}
W.~G. {Dixon}.
\newblock {Mathisson's new mechanics: Its aims and realisation}.
\newblock \emph{Acta Phys. Pol. B Proc. Suppl.}, 1:\penalty0 27, 2008.

\bibitem[{Puetzfeld} and {Obukhov}(2014)]{Puetzfeld:Obukhov:2014:2}
D.~{Puetzfeld} and Yu.~N. {Obukhov}.
\newblock {Equations of motion in metric-affine gravity: a covariant unified
  framework}.
\newblock \emph{Phys. Rev. D}, 90:\penalty0 084034, 2014.

\bibitem[{Dixon}(2015)]{Dixon:2015}
W.~G. {Dixon}.
\newblock {The New Mechanics of Myron Mathisson and its subsequent
  development}.
\newblock \emph{''Equations of Motion in Relativistic Gravity'', D. Puetzfeld
  et. al. (eds.), Fundamental theories of Physics, Springer}, 179:\penalty0 1,
  2015.

\bibitem[{Obukhov} and {Puetzfeld}(2015)]{Obukhov:Puetzfeld:2015:1}
Yu.~N. {Obukhov} and D.~{Puetzfeld}.
\newblock {Multipolar test body equations of motion in generalized gravity
  theories}.
\newblock \emph{''Equations of Motion in Relativistic Gravity'', D. Puetzfeld
  et. al. (eds.), Fundamental theories of Physics, Springer}, 179:\penalty0 67,
  2015.

\bibitem[{Ottewill} and {Wardell}(2011)]{Ottewill:2011}
A.~C. {Ottewill} and B.~{Wardell}.
\newblock {Transport equation approach to calculations of Hadamard Green
  functions and non-coincident DeWitt coefficients}.
\newblock \emph{Phys. Rev. D}, 84:\penalty0 104039, 2011.

\bibitem[{Poisson} et~al.(2011){Poisson}, {Pound}, and
  {Vega}]{Poisson:etal:2011}
E.~{Poisson}, A.~{Pound}, and I.~{Vega}.
\newblock {The motion of point particles in curved spacetime}.
\newblock \emph{Living Reviews in Relativity}, 14\penalty0 (7), 2011.

\bibitem[{Szekeres}(1965)]{Szekeres:1965}
P.~{Szekeres}.
\newblock {The gravitational compass}.
\newblock \emph{J. Math. Phys.}, 6:\penalty0 1387, 1965.

\bibitem[{Pleba\'{n}ski}(1965)]{Plebanski:1965}
J.~{Pleba\'{n}ski}.
\newblock {Conformal geodesic deviations}.
\newblock \emph{Acta Phys. Pol.}, 28:\penalty0 141, 1965.

\bibitem[{Hodgkinson}(1972)]{Hodgkinson:1972}
D.~E. {Hodgkinson}.
\newblock {A modified theory of geodesic deviation}.
\newblock \emph{Gen. Rel. Grav.}, 3:\penalty0 351, 1972.

\bibitem[{Ba\.za\'nski}(1974)]{Bazanski:1974}
S.~L. {Ba\.za\'nski}.
\newblock {The relative energy of test particles in General Relativity}.
\newblock \emph{Nova Acta Leopoldina}, 39:\penalty0 215, 1974.

\bibitem[{Hojman}(1975)]{Hojman:1975}
S.~A. {Hojman}.
\newblock {Electromagnetic and gravitational interactions of a relativistic
  spherical top}.
\newblock \emph{PhD. thesis, Princeton University}, 1975.

\bibitem[{Ba\.za\'nski}(1976)]{Bazanski:1976}
S.~L. {Ba\.za\'nski}.
\newblock {A geometric formulation of the Taylor theorem for curves on affine
  manifolds}.
\newblock \emph{J. Math. Phys. (N.Y.)}, 17:\penalty0 217, 1976.

\bibitem[{Ba\.za\'nski}(1977{\natexlab{a}})]{Bazanski:1977:1}
S.~L. {Ba\.za\'nski}.
\newblock {Kinematics of relative motion of test particles in general
  relativity}.
\newblock \emph{Ann. H. Poin. A}, 27:\penalty0 115, 1977{\natexlab{a}}.

\bibitem[{Ba\.za\'nski}(1977{\natexlab{b}})]{Bazanski:1977:2}
S.~L. {Ba\.za\'nski}.
\newblock {Dynamics of relative motion of test particles in general
  relativity}.
\newblock \emph{Ann. H. Poin. A}, 27:\penalty0 145, 1977{\natexlab{b}}.

\bibitem[{Novello} et~al.(1977){Novello}, {Dami\~ao Soares}, and
  {Salim}]{Novello:etal:1977}
M.~{Novello}, I.~{Dami\~ao Soares}, and J.~M. {Salim}.
\newblock {On Jacobi fields}.
\newblock \emph{Gen. Rel. Grav}, 8:\penalty0 95, 1977.

\bibitem[{Aleksandrov} and {Piragas}(1978)]{Aleksandrov:Piragas:1978}
A.~N. {Aleksandrov} and K.~A. {Piragas}.
\newblock {Geodesic structure: I. Relative dynamics of geodesics}.
\newblock \emph{Theoretical and Mathematical Physics}, 38:\penalty0 48, 1978.

\bibitem[{Manoff}(1979)]{Mannoff:1979}
S.~{Manoff}.
\newblock {Lie derivatives and deviation equations in Riemannian spaces}.
\newblock \emph{Gen. Rel. Grav}, 11:\penalty0 189, 1979.

\bibitem[{Schattner} and {Tr\"umper}(1981)]{Schattner:Truemper:1981}
R.~{Schattner} and M.~{Tr\"umper}.
\newblock {World vectors, Jacobi vectors and Jacobi one-forms on a manifold
  with a linear symmetric connection}.
\newblock \emph{J. Phys. A: Math. Gen.}, 14:\penalty0 2345, 1981.

\bibitem[{Swaminarayan} and {Safko}(1983)]{Swaminarayan:etal:1983}
N.~S. {Swaminarayan} and J.~L. {Safko}.
\newblock {A coordinate-free derivation of a generalized geodesic deviation
  equation}.
\newblock \emph{J. Math. Phys. (N.Y.)}, 24:\penalty0 883, 1983.

\bibitem[{Schutz}(1985)]{Schutz:1985}
B.~{Schutz}.
\newblock {On generalized equations of geodesic deviation}.
\newblock \emph{In: Galaxies, Axisymmetric Systems, and Relativity, Ed. M.A.H.
  MacCallum, Cambridge University Press, Cambridge}, 17:\penalty0 237, 1985.

\bibitem[{Kamran} and {Marck}(1986)]{Kamran:Marck:1986}
N.~{Kamran} and J.-A. {Marck}.
\newblock {Parallel-propagated frame along the geodesics of the metrics
  admitting a Killing-Yano tensor}.
\newblock \emph{J. Math. Phys. (N.Y.)}, 27:\penalty0 1589, 1986.

\bibitem[{Ciufolini}(1986)]{Ciufolini:1986}
I.~{Ciufolini}.
\newblock {Generalized geodesic deviation equation}.
\newblock \emph{Phys. Rev. D}, 34:\penalty0 1014, 1986.

\bibitem[{Ba\.za\'nski} and
  {Kostyukovich}(1987{\natexlab{a}})]{Bazanski:etal:1987:1}
S.~L. {Ba\.za\'nski} and N.~N. {Kostyukovich}.
\newblock {Kinematics of relative motion of charged test particles in general
  relativity. I. The first electromagnetic deviation}.
\newblock \emph{Acta Phys. Pol. B}, 18:\penalty0 601, 1987{\natexlab{a}}.

\bibitem[{Ba\.za\'nski} and
  {Kostyukovich}(1987{\natexlab{b}})]{Bazanski:etal:1987:2}
S.~L. {Ba\.za\'nski} and N.~N. {Kostyukovich}.
\newblock {Kinematics of relative motion of charged test particles in general
  relativity. II. The second electromagnetic deviation}.
\newblock \emph{Acta Phys. Pol. B}, 18:\penalty0 621, 1987{\natexlab{b}}.

\bibitem[{Ba\.za\'nski}(1989)]{Bazanski:1988}
S.~L. {Ba\.za\'nski}.
\newblock {Hamilton-Jacobi formalism for geodesics and geodesic deviations}.
\newblock \emph{J. Math. Phys. (N.Y.)}, 30:\penalty0 1018, 1989.

\bibitem[{Vanzo}(1992)]{Vanzo:1992}
L.~{Vanzo}.
\newblock {A generalization of the equation of geodesic deviation}.
\newblock \emph{Nuovo Cim. B}, 107:\penalty0 771, 1992.

\bibitem[{Roberts}(1996)]{Roberts:1996}
M.~D. {Roberts}.
\newblock {The quantization of geodesic deviation}.
\newblock \emph{Gen. Rel. Grav}, 28:\penalty0 1385, 1996.

\bibitem[{Kerner} et~al.(2001){Kerner}, {van Holten}, and {Colistete
  Jr}]{Kerner:etal:2001}
R.~{Kerner}, J.~W. {van Holten}, and R.~{Colistete Jr}.
\newblock {Relativistic epicycles: another approach to geodesic deviations}.
\newblock \emph{Class. Quantum Grav.}, 18:\penalty0 4725, 2001.

\bibitem[{Manoff}(2001)]{Mannoff:2001}
S.~{Manoff}.
\newblock {Deviation operator and deviation equations over spaces with affine
  connections and metrics}.
\newblock \emph{J. Geom. Phys.}, 39:\penalty0 337, 2001.

\bibitem[{van Holten}(2002)]{Holten:2002}
J.~W. {van Holten}.
\newblock {World-line deviations and epicycles}.
\newblock \emph{Int. J. Mod. Phys. A}, 17:\penalty0 2764, 2002.

\bibitem[{Chicone} and {Mashhoon}(2002)]{Chicone:Mashhoon:2002}
C.~{Chicone} and B.~{Mashhoon}.
\newblock {The generalized Jacobi equation}.
\newblock \emph{Class. Quant. Grav.}, 19:\penalty0 4231, 2002.

\bibitem[{Nieto} et~al.(2003){Nieto}, {Saucedo}, and
  {Villanueva}]{Nieto_etal:2003}
J.~A. {Nieto}, J.~{Saucedo}, and V.M. {Villanueva}.
\newblock {Relativistic top deviation equation and gravitational waves}.
\newblock \emph{Phys. Lett. A}, 312:\penalty0 175, 2003.

\bibitem[{Mohseni}(2004)]{Mohseni:2004}
M.~{Mohseni}.
\newblock {World-line deviation and spinning particles}.
\newblock \emph{Phys. Lett. B}, 587:\penalty0 133, 2004.

\bibitem[{Heydari-Fard} et~al.(2005){Heydari-Fard}, {Mohseni}, and
  {Sepangi}]{Heydari-Fard:etal:2005}
M.~{Heydari-Fard}, M.~{Mohseni}, and H.~R. {Sepangi}.
\newblock {Worldline deviations of charged spinning particles}.
\newblock \emph{Phys. Lett. B}, 626:\penalty0 230, 2005.

\bibitem[{Perlick}(2008)]{Perlick:2008}
V.~{Perlick}.
\newblock {On the generalized Jacobi equation}.
\newblock \emph{Gen. Rel. Grav}, 40:\penalty0 1029, 2008.

\bibitem[{Vines}(2015)]{Vines:2014}
J.~{Vines}.
\newblock {Geodesic deviation at higher orders via covariant bitensors}.
\newblock \emph{Gen. Rel. Grav}, 47:\penalty0 59, 2015.

\bibitem[{Shirokov}(1973)]{Shirokov:1973}
M.~F. {Shirokov}.
\newblock {On one new effect of the Einsteinian theory of gravtiation}.
\newblock \emph{Gen. Rel. Grav}, 4:\penalty0 131, 1973.

\bibitem[{Greenberg}(1974)]{Greenberg:1974}
P.~{Greenberg}.
\newblock {The equation of geodesic deviation in Newtonian theory and the
  oblateness of the Earth}.
\newblock \emph{Nuovo Cim. B}, 24:\penalty0 272, 1974.

\bibitem[{Mashhoon}(1975)]{Mashhoon:1975}
B.~{Mashhoon}.
\newblock {On tidal phenomena in a strong gravtitational field}.
\newblock \emph{Astrophys. J.}, 197:\penalty0 705, 1975.

\bibitem[{Mashhoon}(1977)]{Mashhoon:1977}
B.~{Mashhoon}.
\newblock {Tidal radiation}.
\newblock \emph{Astrophys. J.}, 216:\penalty0 591, 1977.

\bibitem[{Tammelo}(1977)]{Tammelo:1977}
R.~{Tammelo}.
\newblock {Quadrupole Test Particle as a Detector of Gravitational Waves}.
\newblock \emph{Gen. Rel. Grav}, 8:\penalty0 313, 1977.

\bibitem[{Dolan} et~al.(1980){Dolan}, {Choudhury}, and
  {Safko}]{Dolan:etal:1980}
P.~{Dolan}, P.~{Choudhury}, and J.~L. {Safko}.
\newblock {A ``constant of motion'' for the geodesic deviation equation}.
\newblock \emph{J. Austral. Math. Soc. B}, 22:\penalty0 28, 1980.

\bibitem[{Caviglia} et~al.(1982){Caviglia}, {Zordan}, and
  {Salmistraro}]{Caviglia:etal:1982}
G.~{Caviglia}, C.~{Zordan}, and F.~{Salmistraro}.
\newblock {Equation of geodesic deviation and Killing tensors}.
\newblock \emph{Int. J. Theo. Phys.}, 21:\penalty0 391, 1982.

\bibitem[{Fuchs}(1983)]{Fuchs:1983}
H.~{Fuchs}.
\newblock {Solutions of the equations of geodesic deviation for static
  spherical symmetric space-times}.
\newblock \emph{Ann. Phys. (Leipzig)}, 495:\penalty0 231, 1983.

\bibitem[{Audretsch} and {L\"ammerzahl}(1983)]{Audretsch:1983}
J.~{Audretsch} and C.~{L\"ammerzahl}.
\newblock {Local and nonlocal measurements of the Riemann tensor}.
\newblock \emph{Gen. Rel. Grav}, 15:\penalty0 495, 1983.

\bibitem[{Tammelo}(1984)]{Tammelo:1984}
R.~{Tammelo}.
\newblock {On the physical significance of the second geodesic deviation}.
\newblock \emph{Phys. Lett. A}, 106:\penalty0 227, 1984.

\bibitem[{Ba\.za\'nski}(1986)]{Bazanski:1986}
S.~L. {Ba\.za\'nski}.
\newblock {A method of solving the geodesic deviation equations}.
\newblock \emph{Proc. 4th Marcel Grossmann meeting on General Relativity, Ed.
  R. Ruffini, Elsevier (Amsterdam)}, page 1615, 1986.

\bibitem[{Ciufolini} and {Demianski}(1986)]{Ciufolini:Demianski:1986}
I.~{Ciufolini} and M.~{Demianski}.
\newblock {How to measure the curvature of space-time}.
\newblock \emph{Phys. Rev. D}, 34:\penalty0 1018, 1986.

\bibitem[{Ba\.za\'nski} and
  {Kostyukovich}(1987{\natexlab{c}})]{Bazanski:etal:1987:3}
S.~L. {Ba\.za\'nski} and N.~N. {Kostyukovich}.
\newblock {On first integrals of the electromagnetic deviation equations}.
\newblock \emph{Acta Phys. Pol. B}, 18:\penalty0 983, 1987{\natexlab{c}}.

\bibitem[{Ba\.za\'nski} and {Jaranowski}(1989)]{Bazanski:1989}
S.~L. {Ba\.za\'nski} and P.~{Jaranowski}.
\newblock {Geodesic deviation in the Schwarzschild space-time}.
\newblock \emph{J. Math. Phys. (N.Y.)}, 30:\penalty0 1794, 1989.

\bibitem[{Fuchs}(1990{\natexlab{a}})]{Fuchs:1990:1}
H.~{Fuchs}.
\newblock {Paralleltransport and geodesic deviation in static spherically
  symmetric space-times}.
\newblock \emph{Astron. Nach.}, 311:\penalty0 219, 1990{\natexlab{a}}.

\bibitem[{Fuchs}(1990{\natexlab{b}})]{Fuchs:1990:2}
H.~{Fuchs}.
\newblock {Deviation of circular geodesics in static spherically symmetric
  space-times}.
\newblock \emph{Astron. Nach.}, 311:\penalty0 271, 1990{\natexlab{b}}.

\bibitem[{Mohseni} and {Sepangi}(2000)]{Mohseni:2000}
M.~{Mohseni} and H.~R. {Sepangi}.
\newblock {Gravitational waves and spinning test particles}.
\newblock \emph{Class. Quant. Grav.}, 17:\penalty0 4615, 2000.

\bibitem[{Balakin} et~al.(2000){Balakin}, {van Holten}, and
  {Kerner}]{Balakin:etal:2000}
A.~{Balakin}, J.~W. {van Holten}, and R.~{Kerner}.
\newblock {Motions and worldline deviations in Einstein Maxwell theory}.
\newblock \emph{Class. Quantum Grav.}, 17:\penalty0 5009, 2000.

\bibitem[{Colistete} et~al.(2002){Colistete}, {Leygnac}, and
  {Kerner}]{Colistete:etal:2002}
R.~{Colistete}, C.~{Leygnac}, and R.~{Kerner}.
\newblock {Higher-order geodesic deviations applied to the Kerr metric}.
\newblock \emph{Class. Quantum Grav.}, 19:\penalty0 4573, 2002.

\bibitem[{Biesiada}(2003)]{Biesiada:2003}
M.~{Biesiada}.
\newblock {Epicyclic orbital oscillations in Newton's and Einstein's gravity
  from the geodesic deviation equation}.
\newblock \emph{Gen. Rel. Grav.}, 35:\penalty0 1503, 2003.

\bibitem[{Baskaran} and {Grishchuk}(2004)]{Baskaran:Grishchuk:2004}
D.~{Baskaran} and L.~P. {Grishchuk}.
\newblock {Components of the gravitational force in the field of a
  gravitational wave}.
\newblock \emph{Class. Quantum Grav.}, 21:\penalty0 4041, 2004.

\bibitem[{Mullari} and {Tammelo}(2006)]{Mullari:Tammelo:2006}
T.~{Mullari} and R.~{Tammelo}.
\newblock {On the relativistic tidal effects in the second approximation}.
\newblock \emph{Class. Quant. Grav}, 23:\penalty0 4047, 2006.

\bibitem[{Mortazavimanesh} and {Mohseni}(2009)]{Mohseni:2009}
M.~{Mortazavimanesh} and M.~{Mohseni}.
\newblock {Spinning particles in Schwarzschild-de Sitter space-time}.
\newblock \emph{Gen. Rel. Grav.}, 41:\penalty0 2697, 2009.

\bibitem[{Bini} et~al.(2011){Bini}, {Geralico}, and {Jantzen}]{Bini:etal:2011}
D.~{Bini}, A.~{Geralico}, and R.~T. {Jantzen}.
\newblock {Spin-geodesic deviations in Schwarzschild spacetime}.
\newblock \emph{Gen. Rel. Grav}, 43:\penalty0 959, 2011.

\bibitem[{Koekoek} and {van Holten}(2011)]{Koekoek:etal:2011}
G.~{Koekoek} and J.~W. {van Holten}.
\newblock {Geodesic deviations: modeling extreme mass-ratio systems and their
  gravitational waves}.
\newblock \emph{Class. Quantum Grav.}, 28:\penalty0 225022, 2011.

\bibitem[{Petrov}(1969)]{Petrov:1969}
A.~Z. {Petrov}.
\newblock \emph{{Einstein spaces}}.
\newblock Pergamon Press: Oxford, New York, 1969, 1969.

\bibitem[Veblen(1922)]{Veblen:1922}
O.~Veblen.
\newblock {Normal coordinates for the geometry of paths}.
\newblock \emph{Proc. Nat. Acad. Sci. (USA)}, 8:\penalty0 192, 1922.

\bibitem[{Veblen} and {Thomas}(1923)]{Veblen:Thomas:1923}
O.~{Veblen} and T.~Y. {Thomas}.
\newblock {The geometry of paths}.
\newblock \emph{Trans. Amer. Math. Soc.}, 25:\penalty0 551, 1923.

\bibitem[{Thomas}(1934)]{Thomas:1934}
T.~Y. {Thomas}.
\newblock \emph{{The differential invariants of generalized spaces}}.
\newblock Cambridge: University press, 1934.

\bibitem[{Avramidi}(1991)]{Avramidi:1991}
I.~G. {Avramidi}.
\newblock {A covariant technique for the calculation of the one-loop effective
  action}.
\newblock \emph{Nucl. Phys. B}, 355:\penalty0 712, 1991.

\bibitem[{Avramidi}(1995)]{Avramidi:1995}
I.~G. {Avramidi}.
\newblock {Covariant methods for the calculation of the effective action in
  quantum field theory and investigation of higher-derivative quantum gravity}.
\newblock 1995.
\newblock URL \url{hep-th/9510140}.

\end{thebibliography}
\end{document}